\documentclass[iop]{emulateapj}

\usepackage[colorlinks,citecolor=blue, linkcolor=blue]{hyperref}
\usepackage{graphicx}
\usepackage{color,soul}
\usepackage{natbib}
\usepackage{amsmath}
\usepackage{url}

\begin{document}

\title{The Milky Way's Hot Gas Kinematics: Signatures in Current and Future \ion{O}{7} Absorption Line Observations}
\author{Matthew J. Miller, Edmund J. Hodges-Kluck, \& Joel N. Bregman }
\affil{Department of Astronomy, University of Michigan, Ann Arbor, MI 48104, USA}
\email{mjmil@umich.edu, hodgeskl@umich.edu, jbregman@umich.edu}

\begin{abstract}

Detections of $z \approx$ 0 oxygen absorption and emission lines indicate the Milky Way hosts a hot ($\sim 10^6$ K), low-density plasma extending $\gtrsim$50 kpc into the Mily Way's halo. Current X-ray telescopes cannot resolve the line profiles, but the variation of their strengths on the sky constrains the radial gas distribution. Interpreting the \ion{O}{7} K$\alpha$ absorption line strengths has several complications, including optical depth and line of sight velocity effects. Here, we present model absorption line profiles accounting for both of these effects to show the lines can exhibit asymmetric structures and be broader than the intrinsic Doppler width. The line profiles encode the hot gas rotation curve, the net inflow or outflow of hot gas, and the hot gas angular momentum profile. We show how line of sight velocity effects impact the conversion between equivalent width and the column density, and provide modified curves of growth accounting for these effects. As an example, we analyze the LMC sight line pulsar dispersion measure and \ion{O}{7} equivalent width to show the average gas metallicity is $\gtrsim 0.6 Z_{\odot}$ and $b$ $\gtrsim$ 100 km s$^{-1}$. Determining these properties offers valuable insights into the dynamical state of the Milky Way's hot gas, and improves the line strength interpretation. We discuss future strategies to observe these effects with an instrument that has a spectral resolution of about 3000, a goal that is technically possible today.

\end{abstract}


\section{Introduction}
\label{section.introduction}

The observed dynamics of stars and gas in galaxies have played vital roles in interpreting the content, structure, and formation of galaxies.  Baryons respond to the total galactic gravitational field, implying the observed velocity structure probes the underlying dark matter distribution.  Measurements of the \ion{H}{1} rotation curves of late-type galaxies \citep[e.g., ][]{kent87, deblok_etal08}, the stellar velocity dispersions of early type galaxies \citep[e.g., ][]{bernardi_etal05, matkovic_guzman05}, and the orbits of satellite galaxies \citep[e.g., ][]{boylan_kolchin_etal13} have all been crucial in understanding how and why galaxies form the way they do.  Additionally, kinematic observations and simulations show inflows and outflows of multiphase gas are prevalent in the galaxy evolution process \citep[e.g., ][]{kaufmann_etal06, coil_etal11, kacprzak_etal12}.  The latter probe galactic feedback mechanisms while the former probe the accretion of gas onto galaxies that can generate future star formation activity.  All of these methods provide a unique view of the galaxy formation and evolution process, and help shape the view of the Milky Way.

There has been extensive work on the Milky Way's neutral and warm ionized gas kinematics, providing valuable information on the Milky Way's total \ion{H}{1} content and interplay between the Galactic disk and halo.  In the Galactic disk, all-sky maps of 21 cm emission line radiation reveal emission across a range of velocities as a function of Galactic longitude \citep[e.g., ][]{westerhout57}.  Decoding the emission strength as a function of $l$ and $v$ implies the existence of a \ion{H}{1} disk corotating with the stellar disk and with $\sim$10\% its mass along with numerous spiral arm structures \cite[see review by ][]{burton76}.  Ultraviolet absorption line surveys probing gas at $T \sim$10$^5$ K have also revealed absorbers at a wide range of Local Standard of Rest (LSR) velocities.  Absorbers observed at $|v_{LSR}| \lesssim$ 100 km s$^{-1}$ are typically associated with a warm-ionized atmosphere corotating with the disk \citep[e.g., ][]{savage_etal03, bowen_etal08}, while absorbers at $|v_{LSR}| \geqslant$ 100 km s$^{-1}$ are interpreted as discrete clouds or complexes of material, commonly referred to as high velocity clouds \citep[e.g., ][]{wakker_vanwoerden97, sembach_etal03}.  The clouds' origins are still debated, with the leading theories including galactic fountain processes in the Galactic disk, material stripped from orbiting satellite galaxies, or material condensed out of a hotter phase of halo gas \cite[see review by ][]{putman_etal12}.  Regardless of their origins, the clouds' characteristic velocities imply they can supply a substantial amount of neutral gas to the \ion{H}{1} disk and that the clouds interact with the surrounding hot gas material.

The Milky Way and Milky Way-like galaxies also host hot gas distributions between 10$^6$ - 10$^7$ K \citep{spitzer56}, which formed either from supernova explosions in the disk \citep[e.g., ][]{joung_maclow06, hill_etal12} or gas that was shock-heated to the Milky Way's virial temperature as it accreted onto the dark matter halo \citep[e.g., ][]{white_frenk91, cen_ostriker06, fukugita_peebles06}.  Like the cooler gas, the hot gas responds to the Milky Way's gravitational field, implying any kinematic structure is connected with its formation and evolution.  These motions include global rotation from residual or injected angular momentum, turbulence, bulk flows, and net inflow (accretion) or outflow (wind), all of which relate to how the galaxy has formed and evolved.  However, the large spatial extent of the hot gas ($\gtrsim$10 kpc from the Sun) combined with current observational capabilities limit constraints on hot gas dynamics.  This is important since the gas velocity structure can impact hot gas observables and the inferred structure, similar to the \ion{H}{1} analyses discussed above.  

The Milky Way's hot gas distribution is volume-filling on $\gtrsim$10 kpc scales, has density estimates ranging between $10^{-5} - 10^{-3}$ cm$^{-3}$, and a temperature characteristic of the Milky Way's virial temperature, $\approx 2 \times$10$^{6}$ K.  A collisionally ionized plasma at this temperature emits in the soft X-ray band, 0.5 - 2.0 keV, making it the dominant source of the \textit{ROSAT} 3/4 keV background \citep{snowden_etal97}.  All-sky maps from the \textit{ROSAT} All-sky Survey constrained the characteristic densities and temperature of the plasma, but higher resolution spectroscopic observations with current X-ray telescopes provide additional constraints on the gas structure and origin.  

Recent work on hot gas in the Milky Way has relied on \ion{O}{7} and \ion{O}{8} emission and absorption line observations characteristic of a 10$^{6}$ - 10$^{7}$ K plasma \citep{paerels_kahn03}.  The emission lines are observed in $\sim$ 1000 blank fields of view using either sounding rocket experiments \citep{mccammon_etal02}, or the CCDs on board current X-ray telescopes \citep[e.g., ][]{yoshino_etal09, hs12}.  The number of bright X-ray point sources in the soft X-ray band limits the number of absorption line measurements, but they are detected in about 40 active galactic nuclei \citep[e.g., ][]{nicastro_etal02, rasmussen_etal03, wang_etal05, bregman_ld07, yao_wang07, gupta_etal12, miller_bregman13, fang_etal15} and X-ray binary \citep{yao_wang05, hagihara_etal10} spectra.  Interpreting the line strengths has several difficulties: (1) weak detections of \ion{O}{7} K$\beta$ absorption lines suggest some of the lines may not be optically thin \citep[$\tau_o \lesssim$2; ][]{williams_etal05, gupta_etal12, fang_etal15}, (2) the Sun exists within the Local Hot Bubble, which is 100-300 pc in size and has signatures in the soft X-ray band \citep[e.g., ][]{snowden_etal90, snowden_etal93, kuntz_snowden00, lallement_etal03, smith_etal07, welsh_shelton09, smith_etal14}, (3) solar wind charge exchange emission can add significant time-varying emission to individual emission line observations \citep[e.g., ][]{snowden_etal04, koutroumpa_etal07, carter_sembay08, carter_etal11, galeazzi_etal14, hs15}.  Analyses of the line strengths that account for these issues suggest the gas structure is an extended, spherical corona as opposed to a flattened disk-like morphology \citep{fang_etal13, miller_bregman13, miller_bregman15}.

The above studies focused primarily on the line strengths since the lines are unresolved with current X-ray telescopes, however recent data reduction techniques and improved calibration allow the line centroids to be measured with high precision as well.  The emission lines are significantly blended at CCD resolution \citep[EPIC-MOS FWHM $\sim$50 eV $\sim$25,000 km s$^{-1}$ at 0.6 keV; ][]{sembay_etal04}, implying they do not encode hot gas kinematic information.  The absorption lines are also unresolved at current grating resolutions \citep[Reflection Grating Spectrometer FWHM $\sim$2 eV$\sim$1000 km s$^{-1}$ at 0.6 keV; ][]{denherder_etal03}, but the higher resolution compared to the CCDs yields accurate equivalent width and line centroid measurements in high signal-to-noise ratio targets.  In particular, \cite{hodgeskl_etal16} utilized the improved calibration for sun angle and heliocentric motion in \textit{XMM-Newton} RGS observations to significantly improve the precision for line centroid measurements.  This updated calibration resulted in a sample of 37 \ion{O}{7} absorption line centroids from archival \textit{XMM-Newton} RGS data with uncertainties ranging from $\approx$25-400 km s$^{-1}$.  The measured values are inconsistent with a stationary hot gas halo, and suggest the hot rotates in the same direction as the disk.  This analysis highlights that high resolution absorption line observations are useful probes of the Milky Way's hot gas kinematics.  

Here, we present model absorption line profiles for different hot gas velocity structures and analyze their effects on current and future high resolution absorption line observations.  The modeling is analogous to the long-established work on the Milky Way's \ion{H}{1} structure and kinematics discussed above, however the hot gas system we consider is extended and spherical out to the Milky Way's virial radius.  For velocity profiles, we explore the effects of gas rotation in the same direction as the disk and global inflows/outflows of gas.  We show that the absorption lines can exhibit significant asymmetries, broadening beyond the Doppler width, and different line centers depending on the underlying velocity structure.  These model line centroids complement the work discussed above, but observing the profile shapes requires a higher resolution X-ray spectrograph.  We discuss the instrument requirements to observe such effects and how future observations will provide additional constraints on the Milky Way's hot gas kinematics and baryon content \citep[see discussion by ][]{bregman_etal15}.

In addition to the line profile calculations, we show how optical depth and kinematic effects impact ongoing work on the Milky Way's hot gas structure.  Accounting for velocity flows in the halo can broaden the total absorption line profile in velocity space, which impacts the conversion between observed equivalent widths and the inferred column densities.  We present curves of growth that account for these velocity effects along with the plasma optical depth, providing more accurate inferences for the gas density structure.  

The rest of the paper is structured as follows.  In Section ~\ref{section.calculations}, we describe our line profile calculation, including a discussion on the inferred hot gas density and velocity structures.  This also includes the altered conversions between equivalent widths and column densities.  Section ~\ref{section.discussion} includes a discussion and summary of our results, with additional implications for future X-ray missions.

\section{Calculations of Line Shapes and Equivalent Widths}
\label{section.calculations}

The calculation and interpretation of absorption line shapes is analogous to the Galactic \ion{H}{1} distribution.  In these studies, resolved \ion{H}{1} 21 cm emission line measurements in velocity space constrain the neutral hydrogen distribution and kinematics in the Galactic disk \citep[e.g., ][]{kalberla_etal07}.  These methods involve analyzing the amount of emission received at different velocities and different Galactic longitudes.  Here, we make similar predictions for the hot gas absorption line profiles, but assuming different underlying hot gas density and velocity profiles compared to the neutral hydrogen distribution.  The primary difference here is the assumption that the hot gas is a volume-filled distribution of material extending to the Milky Way's virial radius as opposed to only being confined to the Galactic disk.  Similarly, we explore different large-scale velocity profiles for the hot gas that deviate from simple corotating motion with the disk.  These factors imply that different density and velocity profiles produce unique line shapes and line center shifts for different $l, b$ coordinates.  We explore these effects and estimate their impact on current and future absorption line measurements.

\subsection{Model Assumptions}
\label{subsection.model_assumptions}

We make several assumptions for the hot gas density, temperature, and metallicity distribution based on previous observational and theoretical analyses.  Most of these assumptions are based on the results from \cite{miller_bregman13, miller_bregman15}, and we refer the reader to these studies for a more detailed description.  Here, we outline the most important assumptions and any caveats in terms of how they could affect our absorption line profile calculations.

We assume the hot gas density distribution follows a modified spherical $\beta$-model extending to the Milky Way's virial radius.  The $\beta$-model has the following functional form:

	\begin{equation}
	 \label{eq.beta_model}
	 n(r) = {n_o}({1 + ({r}/{r_c})^2})^{-{3\beta}/{2}},
	\end{equation}

\noindent where $n_o$ is the core density, $r_c$ is the core radius, and $\beta$ controls the slope at $r \gg r_c$.  The Milky Way's expected $r_c$ is ($\lesssim$5 kpc), a region that is not well sampled by absorption or emission line observations.  To account for this, \cite{miller_bregman13} defined a modified $\beta$-model in the limit where $r \gg r_c$:

	\begin{equation}
	 \label{eq.beta_model_approx}
	 n(r) \approx \frac{n_or_c^{3\beta}}{r^{3\beta}},
	\end{equation}

\noindent where $n_or_c^{3\beta}$ is the density normalization and -3$\beta$ is the slope.  This density profile is effectively a power law describing the Milky Way's hot gas distribution.  Unless otherwise stated, we assume density parameters of $n_or_c^{3\beta}$ = $1.3 \times 10^{-2}$ cm$^{-3}$ kpc$^{3\beta}$ and $\beta$ = 0.5 based on results from \cite{miller_bregman15}.

This model, although simple, is not arbitrarily chosen, and has successfully reproduced several Milky Way hot gas observables.  Most recent studies on the Milky Way's hot gas structure follow a general methodology.  One assumes one of two types of underlying density distributions: an exponential disk with scale height of 5-10 kpc, or a lower-density, spherical model extending to the virial radius.  Given a model choice, one compares model observations to measured observations and determines if the model is consistent with the data.  The primary observables of the Milky Way's hot gas are \ion{O}{7} and \ion{O}{8} absorption and emission line strengths \citep[e.g., ][]{paerels_kahn03, rasmussen_etal03}.  Several analyses argue that an exponential disk-like morphology can reproduce the line strengths for \textit{individual} sight lines \citep{yao_wang07, hagihara_etal10}.  However, independent analyses on all-sky samples of absorption and emission line strengths suggest a density model like Equation~\ref{eq.beta_model_approx} reproduces the global line strength trends better than a disk-like morphology \citep{miller_bregman13, miller_bregman15, hodgeskl_etal16}.  Additional work by \cite{fang_etal13} explored more sophisticated spherical, extended density models (an adiabatic profile and a cuspy NFW profile) and an exponential disk model to compare with other observed constraints, such as the residual pulsar dispersion measure toward the Large Magellanic Cloud (LMC) \citep{anderson_bregman10} and ram-pressure stripping of dwarf spheroidal galaxies \citep[e.g., ][]{grcevich_putman09}.  They found the spherical, extended models were consistent with these constraints, while the exponential disk model was not.  Therefore, an extended, spherical density morphology appears to be consistent with several observed hot gas properties, and a simple power law can reproduce how the observed line strengths vary across the sky.

The $\beta$-model is also used for fitting the observed X-ray surface brightness profiles around nearby galaxies.  This has historically been done for $\approx$50 early-type galaxies, with fitted $\beta$ values ranging between 0.4 - 1.0 for a typical early-type galaxy \citep{forman_etal85, osullivan_etal03}.  More recently, there have been detections of diffuse X-ray halos around massive ($\sim$10 times larger than the Milky Way) late-type galaxies within $\sim$70 kpc \citep{anderson_bregman11, dai_etal12, bogdan_etal13b, bogdan_etal13a, anderson_etal16}.  Although there are only four late-type galaxies where these coronae are detected, their fitted $\beta$ values are also comparable to the results from early-type galaxy surveys.  This further motivates our initial choice to use a $\beta$ model as our underlying Milky Way hot gas density profile.

We also assume that the hot gas temperature, oxygen abundance, and metallicity are constant with radius.  Observationally, the Milky Way's hot gas temperature is inferred from a combination of fitting soft X-ray background (SXRB) spectra with thermal plasma models, or comparing the \ion{O}{7} to \ion{O}{8} absorption line ratio in individual sight lines where both lines are detected.  Both methods indicate that the hot gas temperature is $\approx 2 \times 10^6$ K, with the latter method providing the most current observational constraints.  \cite{hs13} fit the 0.5 - 2.0 keV band of 110 SXRB spectra with multiple thermal APEC plasma models to measure the Milky Way's hot halo emission measure and temperature distribution.  The found the halo gas temperature shows little variation across the sky with a median value of $2.22 \times 10^6$ K and interquartile region of $0.63 \times 10^6$ K.  The fact that this value is consistent across the sky implies the gas is close to isothermal.  Therefore, we assume a constant halo gas temperature of $2 \times 10^{6}$ K, resulting in an \ion{O}{7} ion fraction of 0.5 \citep{sutherland_dopita93}.

There is less certainty when considering the hot gas metallicity and oxygen abundance since there are no direct observational constraints on these values.  The reported solar oxygen abundance relative to hydrogen has varied in the literature, but we assume a value of 5.49 $\times$ 10$^{-4}$ \cite{holweger_01}.  Galaxy evolution simulations suggest halo gas metallicities can range between $\approx$0.05 - 2 $Z_{\odot}$ depending on the stellar feedback prescription, with a typical value of $\approx$0.3$Z_{\odot}$ \citep{toft_etal02, cen_ostriker06, marinacci_etal14}.  This is consistent with the model-dependent constraint \cite{miller_bregman15} placed on the hot gas metallicity in order to not overproduce the pulsar dispersion measure in the LMC direction, $Z \geq 0.3Z_{\odot}$.  \cite{miller_bregman15} also estimated a model-dependent hot gas metallicity profile by comparing their emission line fitting results (sensitive to $n_e n_{ion}$) with absorption line results (sensitive to $n_{ion}$) from \cite{miller_bregman13}.  They found evidence for a shallow metallicity gradient of $Z \propto r^{-0.2}$ with $Z = 0.26Z_{\odot}$ at $r$ = 10 kpc, although the uncertainty in the relation was consistent with a constant gas metallicity.  Thus, we assume a constant metallicity value of 0.3$Z_{\odot}$, which still results in $\gtrsim$50\% of the absorption coming from $\gtrsim$5 kpc of the Sun and $\gtrsim$90\% coming from $\lesssim$50 kpc \citep{hodgeskl_etal16}.

It is worth noting that these hot gas distribution assumptions are all broadly consistent with the theoretical prediction that the hot gas formed via an accretion shock and is in hydrostatic equilibrium with the Milky Way's dark matter potential well.  Our density parametrization of a power law is the simplest functional form that can remain in hydrostatic equilibrium with the dark matter potential well at large radii, with $\beta$ = 0.5 implying the gas should be close to hydrostatic equilibrium.  Given this type of distribution and formation mechanism, one expects the gas temperature to be close to the Galactic virial temperature of 2 $\times$ 10$^6$ K.  This prediction is consistent with the observed, and therefore assumed hot gas temperature.  The global hot gas distribution adopted here has a reasonable theoretical basis, and is also consistent with multiple types of observations.

Finally, our calculations depend on several Milky Way parameters that have been measured from different techniques.  For the Milky Way's size and geometry, we assume a solar distance to the Galactic center of $R_{\odot}$ = 8.34 kpc, a solar rotation velocity around the Galactic center of $v_{\odot}$ = 240 km s$^{-1}$ \citep{reid_etal14}, and an average virial radius of $r_{v}$ = 250 kpc \citep{klypin_etal02, loeb_etal05, shattow_loeb09}.  The former do not affect our overall line shapes, but act as a net shift in wavelength/velocity space for the line profiles.  We extrapolate our model profiles and calculations to $r_v$, however gas within $\approx$50 kpc of the Sun dominates our column densities and equivalent widths \citep{miller_bregman13, hodgeskl_etal16}.  This implies that our choice for $r_v$ has a negligible impact on our results, and that our model profiles are explicitly valid within $\approx$50 kpc.

\subsection{Velocity Profile and Line Profile Calculation}
\label{subsection.line_profile}

We consider bulk rotation motion, radial inflow/outflow, and flows perpendicular to the Galactic disk for our velocity profiles.  A net rotational flow (denoted as the $\hat{\phi}$ direction henceforth) is expected if the gas has residual angular momentum from the Milky Way's formation.  The radial and perpendicular flows ($\hat{r}$ and $\hat{z}$ directions) are considered as a net accretion of material onto the Galactic disk or a net ejection of material from the Galactic disk.  We assume constant flow velocities as a function of $R$, $z$, or $r$, depending on the flow type, where $R$ is the distance from the rotation axis, $z$ is the distance perpendicular to the Galactic plane, and $r$ is the galactocentric radial distance.  These are represented by the following equations:

	\begin{equation}
	 \label{eq.v_phi}
	 v_{\phi}(R) = v_{\phi} \hat{\phi} = constant, 
	\end{equation}\\[-34pt]

	\begin{equation}
	 \label{eq.v_r}
	 v_{r}(r) = v_{r} \hat{r} = constant, 
	\end{equation}\\[-34pt]

	\begin{equation}
	 \label{eq.v_z}
	 v_{z}(z) = v_{z} \hat{z} = constant, 
	\end{equation}

\noindent where $v_{\phi}$ follows a flat rotation curve similar to the disk, and $v_{r}$/$v_{z}$ are net flows in the $\hat{r}$/$\hat{z}$ directions respectively.  Therefore, positive $v_{r}$/$v_{z}$ values assume flows away from the Galactic center ($v_{r}$) or Galactic plane ($v_{z}$).  We also consider models with a constant mass inflow/outflow rate by allowing $v_{r}(r)$ to not be a constant.  The mass flux depends on the density and radial velocity, and is represented by the following equation:

	\begin{equation}
	 \label{eq.mdot_r}
	 \dot{M_{r}}(r) = 4 \pi r^2 \rho (r) v_{r} (r) \hat{r} = constant, 
	\end{equation}

\noindent where $\rho (r)$ is the mass density as a function of radius.  If $\rho (r) \propto n(r) \propto r^{-3\beta}$, this implies $v_{r} (r) \propto r^{3\beta - 2}$ in order for $\dot{M_r}$ to be constant.  

\begin{figure*}[t!]
\centering
\includegraphics[width = 1.0\textwidth, keepaspectratio=true]{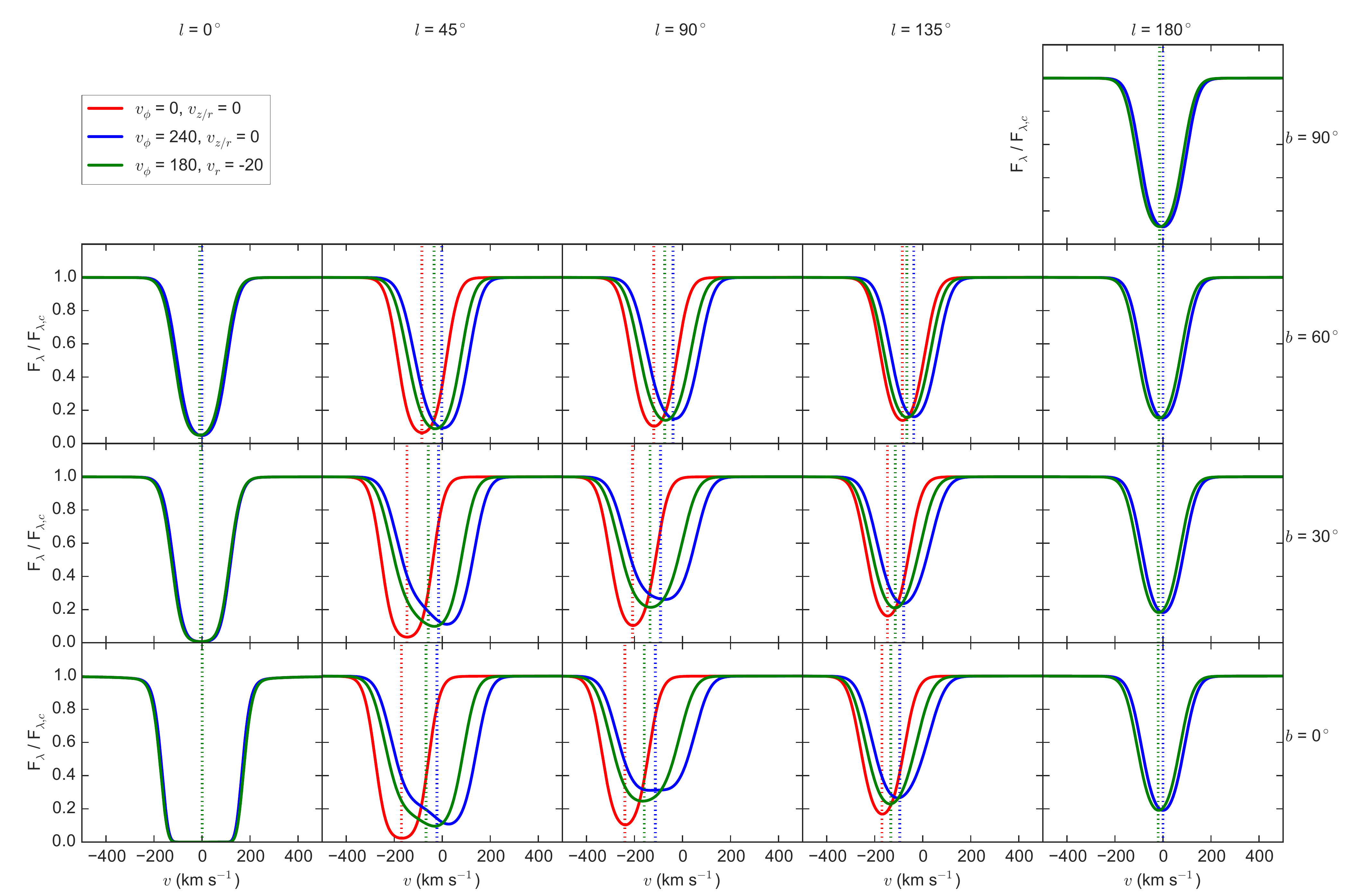}
\caption{Model absorption line profiles, normalized to the continuum, for different positions on the sky and velocity profiles.  Each model spectrum assumes the gas density follows a modified $\beta$-model (Equation~\ref{eq.beta_model_approx} with $\beta$ = 0.5) and that the gas is either stationary (red), corotating with the disk (blue), or lagging behind the disk with a modest radial inflow velocity (green).  The model profiles exhibit asymmetric shapes and varying line centroids (vertical dotted lines) depending on the observed position and underlying velocity profile.
}
\label{figure.line_grid}
\end{figure*}

Optical depth effects and the assumed absorption line Doppler width play an important role when inferring column densities from measured equivalent widths.  The observed absorption lines are never broader than the instrumental line widths of current X-ray spectrographs (the Chandra Low-energy Transmission Grating has a FWHM of $\approx$0.05 \AA\ ).  This sets an upper limit on the Doppler width of $\lesssim$550 km s$^{-1}$.  Given these spectral resolutions, it becomes difficult to quantify deviations from an optically thin plasma.  There have been several detections of the \ion{O}{7} K$\beta$ transition along with the \ion{O}{7} K$\alpha$ line, suggesting Doppler widths ranging from the thermal width ($\approx$45 km s$^{-1}$ at $2 \times 10^6$ K) to $\approx$150 km s$^{-1}$ \citep{williams_etal05, gupta_etal12, fang_etal15}.  Galaxy formation and evolution simulations predict a similar range of $b$ values with a characteristic value of 85 km s$^{-1}$ \citep{cen12}.  The additional turbulent velocity can arise from a variety of sources, including turbulent mixing layers, satellite galaxy orbits, bulk flows of material in and out of the disk, etc. \citep[e.g., ][]{kwak_shelton10, cen12, hill_etal12}.  We adopt a $b$ value of 85 km s$^{-1}$ for our initial line profile calculations, but explore how the Doppler width compares with different velocity flows.

Given a density profile, velocity profile, and Doppler width for the absorbing medium, we calculate model absorption line profiles for different sight lines in the Galaxy.  For a given $l, b$ coordinate, we divide the line of sight into cells with densities and Galactic standard of rest (GSR) velocities dependent on the cells' locations in the Galaxy.  The coordinate transformations between the line of sight position, $s$, and Galactic coordinates is:\\[-.5cm]

	\begin{equation}
		R^2 = R_{\odot}^2 + s^2\cos(b)^2 - 2sR_{\odot}\cos(b)\cos(l)
	\end{equation}\\[-1.2cm]
	
	\begin{equation}
		z^2 = s^2\sin(b)^2
	\end{equation}\\[-1.2cm]
	
	\begin{equation}
		r^2 = R^2 + z^2
	\end{equation}\\[-.8cm]

\noindent The density in each cell comes from the cells' $r$ positions and using Equation (\ref{eq.beta_model_approx}).  The conversions between GSR and LSR velocities for our velocity models are represented by the following equations:\\[-.8cm]

	\begin{equation}
	 \label{eq.v_phi_los}
	 v_{s,\phi}(s) = \left (\frac{v_{\phi}}{R(s)} - \frac{v_{\odot}}{R_{\odot}} \right) R_{\odot} \sin(l)\cos(b),
	\end{equation}\\[-1.1cm]

	\begin{gather}
	\begin{split}
	 \label{eq.v_r_los}
	 v_{s,r}(s) &= \frac{v_{r}}{r(s)} [ \sin(|b|)z(s) 
	 \\ 
	 &+ (s\cos(b) - R_{\odot}\cos(l))\cos(b)],
	\end{split}
	\end{gather}\\[-1.1cm]

	\begin{equation}
	 \label{eq.v_z_los}
	 v_{s,z}(s) = v_{z}\sin(|b|),
	\end{equation}

\noindent where $R(s)$, $z(s)$, and $r(s)$ are the Galactic coordinates along the sight line $s$, $v_{\phi/r/z}$ are the GSR flow velocities, and $v_{s, \phi/r/z}$ are the LSR velocities for each cell.  We generate a Voigt profile for each cell that is shifted to the cell's LSR velocity and weighted by the density, or optical depth, in the cell.  The cell line center optical depth is defined as:

	\begin{equation}
    \tau_o = .015 \times n(s) f \lambda b^{-1} ds,
	\end{equation}

\begin{figure}[!h]
\centering
\includegraphics[width = .5\textwidth, keepaspectratio=true]{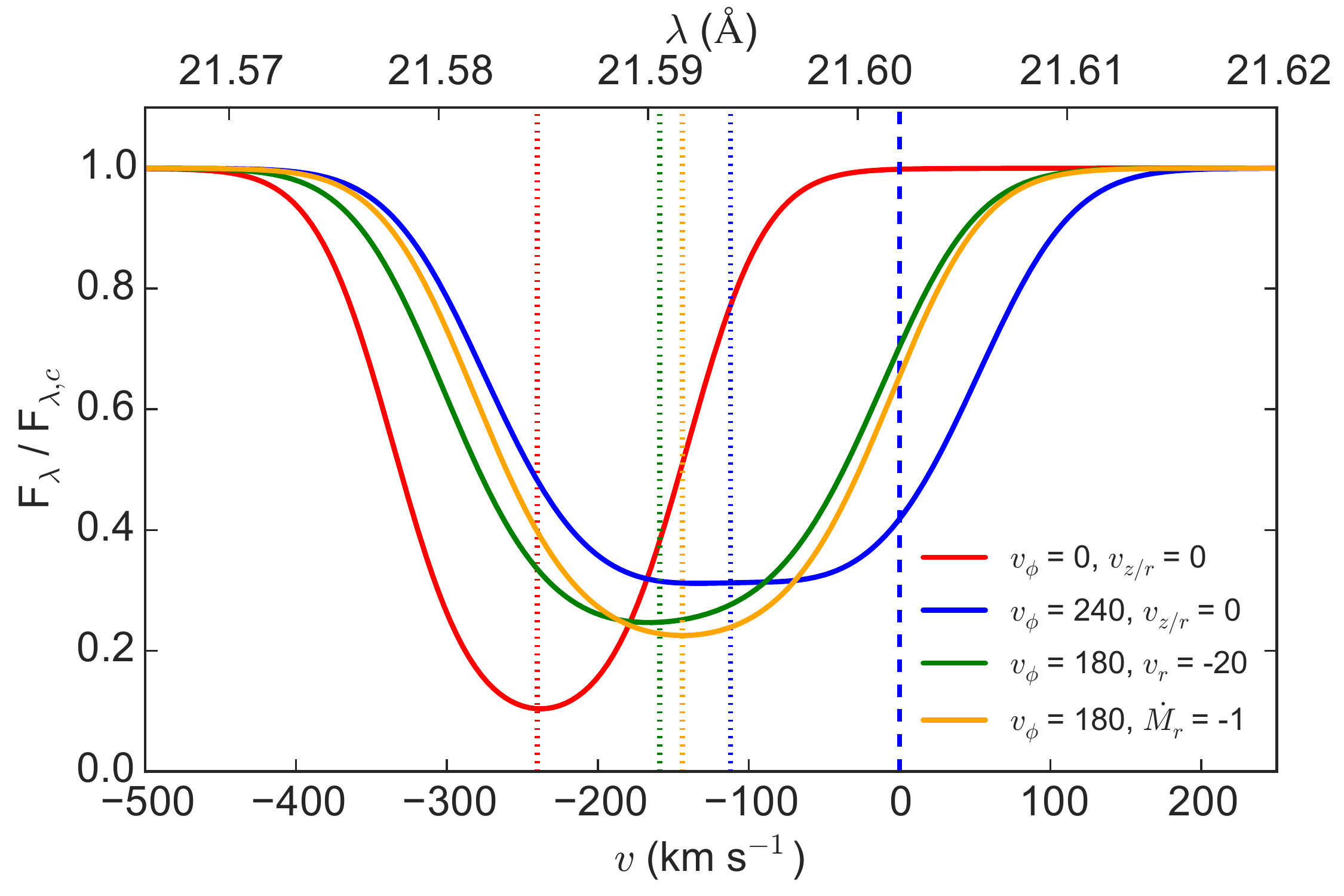}
\caption{Same model spectra as Figure~\ref{figure.line_grid}, except focusing on an $l,b$ = 90$\arcdeg$, 0$\arcdeg$ sight line.  The velocity effects are apparent, where the corotating line profile (blue) is much broader than the stationary line profile (red).  The additional orange line is for a constant radial mass flux model with $\dot{M_{r}}$ = -1 $M_{\odot}$ yr$^{-1}$, which has minor differences from the constant $v_r$ model.  The dashed vertical blue line represents the tangent point velocity for this sight line (0 km s$^{-1}$), although there is still absorption at $v > $ 0 km s$^{-1}$ due to the inferred $b$ value of 85 km s$^{-1}$.  
}
\label{figure.line_detail}
\end{figure}


\noindent where $n(s)$ is the number density of absorbers in the cell, $f$ is the transition oscillator strength ($f$ = .6945 for the \ion{O}{7} K$\alpha$ transition), $\lambda$ is the transition wavelength in centimeters, $b$ is the Doppler width of the plasma in cm s$^{-1}$, .015 is a constant with units cm$^2$ s$^{-1}$, and $ds$ is the cell path length in centimeters.  The total model line optical depth, $\tau _{v}$ or $\tau _{\lambda}$, is the sum of each cell's line profile out to the Milky Way's virial radius with the resultant absorption line profile following the usual definition:

	\begin{equation}
    \frac{F_{v/ \lambda}}{F_{v/ \lambda, c}} = e^{-\tau_{v/ \lambda}},
	\end{equation}

\noindent where $F_{v/ \lambda}$ is the source flux in velocity or wavelength space and $F_{v/ \lambda, c}$ is the continuum flux.

\begin{figure*}[!ht]
\centering
\includegraphics[width = 1.\textwidth, keepaspectratio=true]{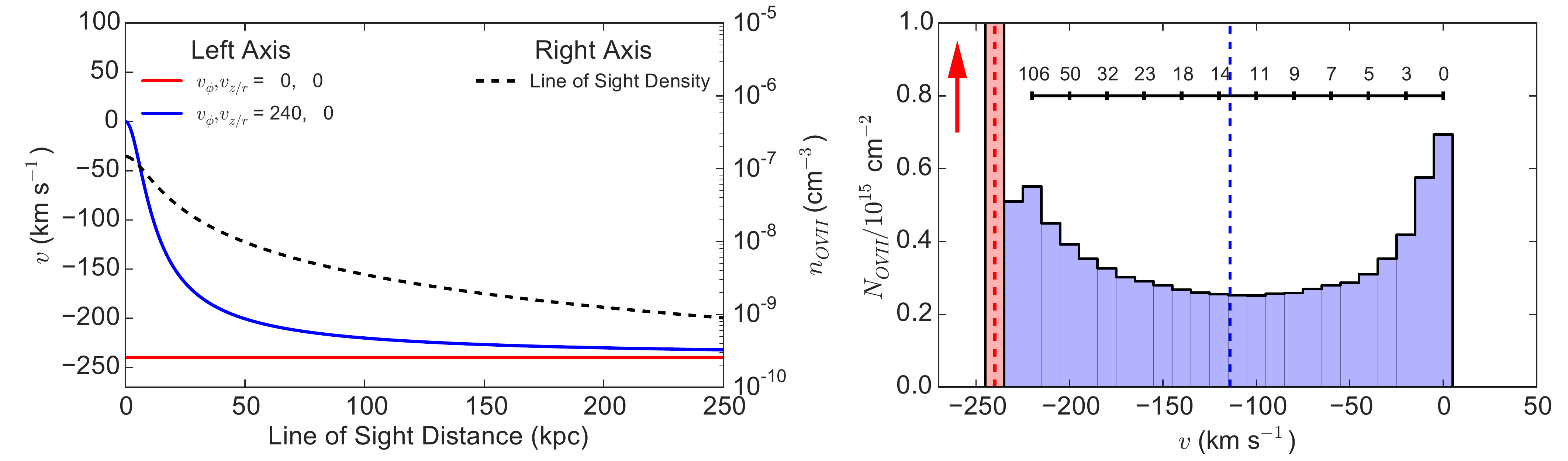}
\caption{The $l,b$ = 90$\arcdeg$, 0$\arcdeg$ line of sight density (dashed line) and velocity (solid lines) profiles (\textit{left}) and \ion{O}{7} column density in 10 km s$^{-1}$ bins along the line of sight (\textit{right}).  In the stationary case (red), all of the gas is at the reflex velocity of the Sun around the Galactic center, -240 km s$^{-1}$ (the column density bin goes above the plot limits to a value of 8.4 in order to emphasize the corotating velocity effects).  In the corotating case (blue), the column density is spread out in velocity space, with the black horizontal scale indicating the median line of sight distance in kpc for every other velocity bin.  Material near the Sun produces $\approx$0 km s$^{-1}$ absorption while material further in the halo produces $\approx$-200 km s$^{-1}$ absorption.
}
\label{figure.los_structure}
\end{figure*}

The resultant model line profiles include asymmetric line shapes, widths broader than the plasma Doppler width, and significant line center shifts.  Figure~\ref{figure.line_grid} shows our absorption line calculations for a grid of 0$^{\circ}$ $\leq$ $l$ $\leq$ 180$^{\circ}$ and 0$^{\circ}$ $\leq$ $b$ $\leq$ 90$^{\circ}$ coordinates.  The velocity profiles include a stationary hot gas model ($v_{\phi}$ = $v_{r/z}$ = 0 km s$^{-1}$), a hot gas distribution corotating with the disk ($v_{\phi}$ = 240 km s$^{-1}$, $v_{r/z}$ = 0 km s$^{-1}$), and a profile similar to the best-fit model from \cite{hodgeskl_etal16} that best reproduces a sample of observed line centroids ($v_{\phi}$ = 180 km s$^{-1}$, $v_{r}$ = -20 km s$^{-1}$).  One immediately sees the line profiles can be highly asymmetric or broadened beyond the assumed 85 km s$^{-1}$ for observations near $l \approx 90^{\circ}$ and for $b \lesssim 30^{\circ}$.  Figure~\ref{figure.line_detail} shows a detailed plot of the model profiles toward the $l, b$ = 90$^{\circ}$, 0$^{\circ}$ direction, where the velocity effects will be strongest.  This figure includes a constant $\dot{M_{r}}$ model line profile with $v_{\phi}$ = 180 km s$^{-1}$, $\dot{M_{r}}$ = -1 $M_{\odot}$ yr$^{-1}$.  There are minor differences between the line shapes and centroids ($\approx$15 km s$^{-1}$) for this model and the $v_{r}$ = -20 km s$^{-1}$ model, so we only consider the constant $v_{r}$ model for the rest of the analysis.

\begin{figure*}[!ht]
\centering
\begin{tabular}{l l}
\includegraphics[width = 0.45\textwidth]{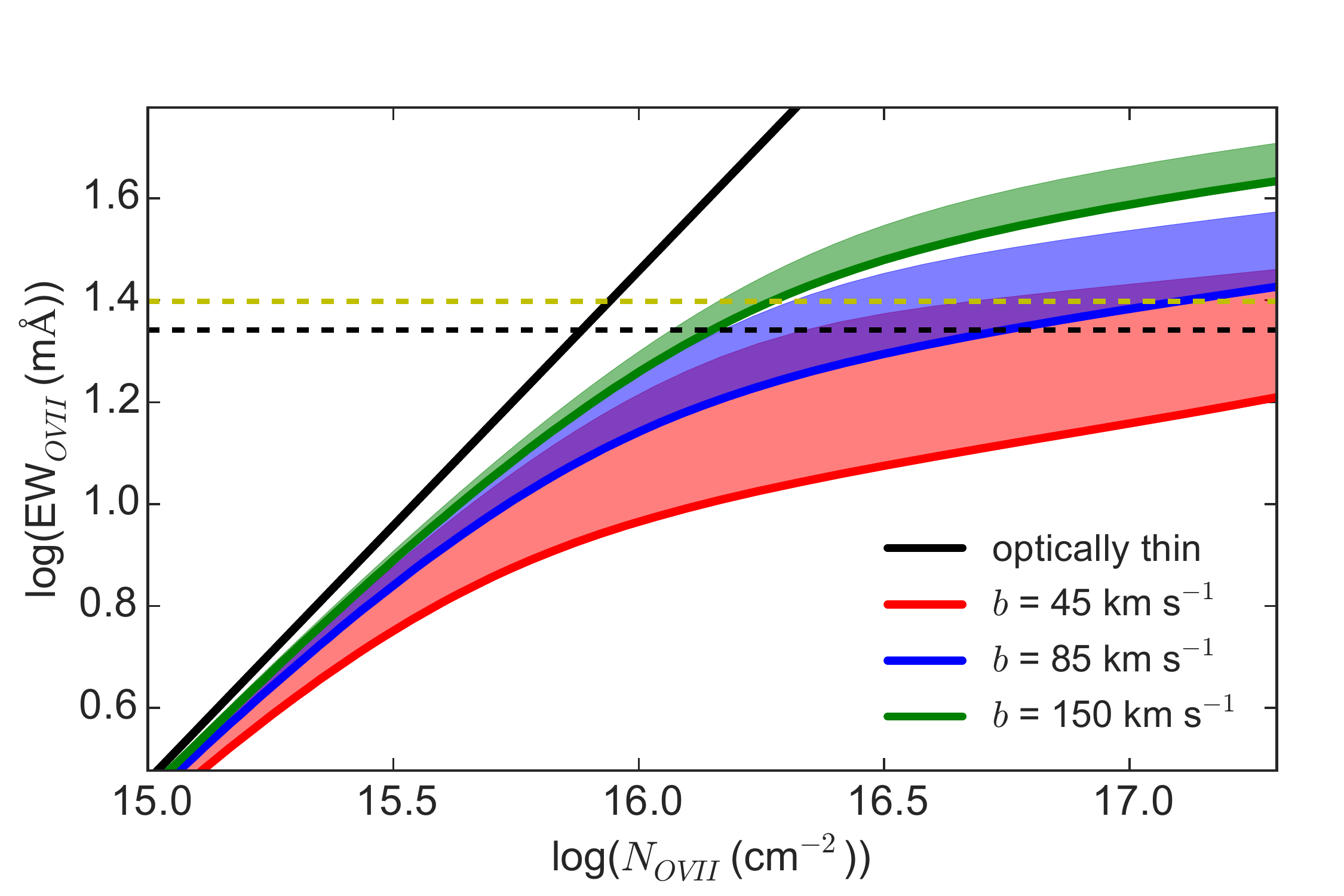}
\includegraphics[width = 0.45\textwidth]{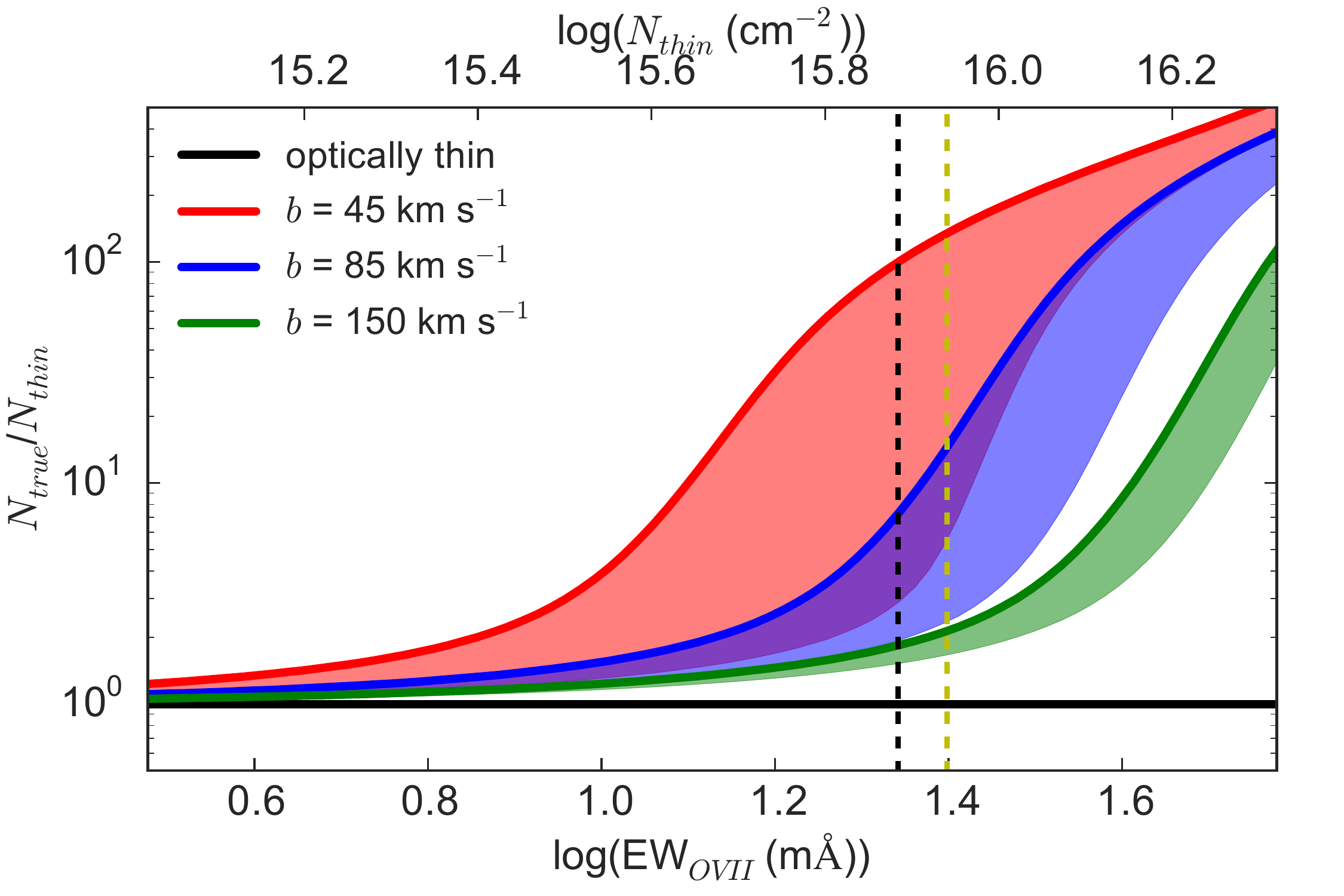}
\end{tabular}
\caption{Conversions between the observed equivalent widths and column densities for different Doppler values and with velocity effects along an $l,b$ = 90$\arcdeg$, 0$\arcdeg$ line of sight.  The \textit{left} panel shows curves of growth, where the colored solid lines represent normal growth curves for different $b$ values (or equivalently a stationary halo), and the upper boundaries of the shaded regions represent the growth curves assuming a corotating velocity profile.  The \textit{right} panel is the same as the \textit{left} panel, but with the column densities normalized to the optically thin values.  The dashed lines represent the median equivalent widths from the \cite{miller_bregman13} (black) and \cite{fang_etal15} (yellow) samples, indicating the optical depth and velocity effects can be substantial.
}
\label{figure.cog_corr}
\end{figure*}

There are several apparent trends seen in these figures.  If the halo gas is stationary, all of the absorption occurs at the reflex motion of the Sun's orbit around the Galactic center, $v_s$ = $-v_{\odot} \sin(l) \cos(b)$.  If instead the hot gas corotates with the disk, the absorption can be spread out between $\pm v_{\odot}$ depending on the direction observed (Figure~\ref{figure.los_structure}).  The resultant density-weighted LSR velocities create the asymmetric line profiles and significantly broader lines compared to the stationary plasma results.  For example, the corotating model line width (defined as the full width at half the line depth) for the $l, b$ = 90$^{\circ}$, 0$^{\circ}$ direction in Figure~\ref{figure.line_detail} is 326 km s$^{-1}$ compared to 200 km s$^{-1}$ for the stationary model, a 63 \% increase.  Furthermore, the absorption line centroids depend on the underlying velocity profile (vertical dotted lines in Figures~\ref{figure.line_grid} and \ref{figure.line_detail}).  The stationary plasma halo produces the strongest deviations from the rest wavelength, or 0 km s$^{-1}$, since all of the absorption is at the same velocity.  If the gas is corotating with the disk, one expects weaker line shifts since dense gas near the Galactic disk produces $v_s \approx$0 km s$^{-1}$ absorption, while gas futher in the halo produces absorption at $|v_s| \gtrsim$100 km s$^{-1}$.  Flows in the $r$ or $z$ directions produce similar net velocity shifts, although flows in these directions have stronger effects near the Galactic center, anti-center, and poles.  These effects imply that the hot gas kinematics produce observable signatures on the absorption line centroids, and alter the inferred conversion between measured equivalent width and column density.

\subsection{Model Line Strengths}
\label{subsection.equivalent_widths}

These line of sight velocity effects have important consequences when analyzing the absorption line strengths.  Specifically, these effects impact observable quantities (equivalent widths, resolved line profiles, etc.) and how we infer the gas column density.  This is not a new concept in the broad field of absorption line analysis, but these large scale velocity flows have never been considered when analyzing the Milky Way's \ion{O}{7} absorbers.  Here, we discuss how these effects impact current line strength interpretations when the lines are unresolved and potential future analyses if the lines can be partially resolved.

Modeling the line of sight velocity flows and corresponding line profiles leads to a more accurate conversion between measured absorption line equivalent widths and the inferred column densities.  The measured equivalent widths are currently the best way to measure the absorption line strengths since the lines are unresolved by current X-ray spectrographs.  The inferred column densities are usually determined from a curve of growth analysis with various assumptions for the plasma optical depth.  Velocity flows alter this conversion by spreading out the absorption in velocity space, producing larger equivalent widths for a given column density.  To quantify the strongest velocity corrections, we generate model line spectra in the $l, b$ = 90$^{\circ}$, 0$^{\circ}$ direction, for a range of column densities and Doppler $b$ values, and for a corotating halo.  We vary the column density by changing the halo normalization parameter in Equation~\ref{eq.beta_model_approx}, thus keeping the gas distribution the same with $\beta$ = 0.5.  We create curves of growth with velocity effects by calculating equivalent widths for each line profile/column density.  Figure~\ref{figure.cog_corr} shows these new growth curves, where the solid lines represent traditional curve of growth calculations with no velocity effects and the opposite edges of the shaded regions represent our curves of growth with rotational velocity effects.  Therefore, the shaded regions represent the range of possible velocity corrections, depending on the true halo velocity structure.  

These velocity effects have a significant impact on the inferred column densities, with potentially large deviations from optically thin values.  Here, we examine corrections to the inferred optically thin column densities with and without considering velocity effects and for different $b$ values.  The right hand panel of Figure~\ref{figure.cog_corr} rescales the curves of growth in the left hand panel into the ratio between the true and optically thin column density for a given equivalent width.  The Doppler $b$ value plays a critical role when inferring the correct column density for a given measured equivalent width, where corrections without velocity effects range between $\approx$50\% to over an order of magnitude for $b$ between 45 and 150 km s$^{-1}$ and a typical equivalent width measurement of 20 m\AA\ .  The velocity effects discussed above mitigate these differences, but still cause significant deviations from optically thin values.  For example, the ratio between the true and optically thin column ranges from a factor of $\approx$2 if the halo has is corotating to a factor of $\approx$5 without velocity effects for $b$ = 85 km s$^{-1}$ and a 20 m\AA\ equivalent width.  This implies that velocity effects result in \textit{smaller} column densities than would be inferred from a traditional curve of growth conversion.  These results are useful in the current limit where the lines are unresolved, but there are alternative approaches if the lines are partially resolved.

\begin{figure*}[!ht]
\centering
\includegraphics[width = 1.\textwidth]{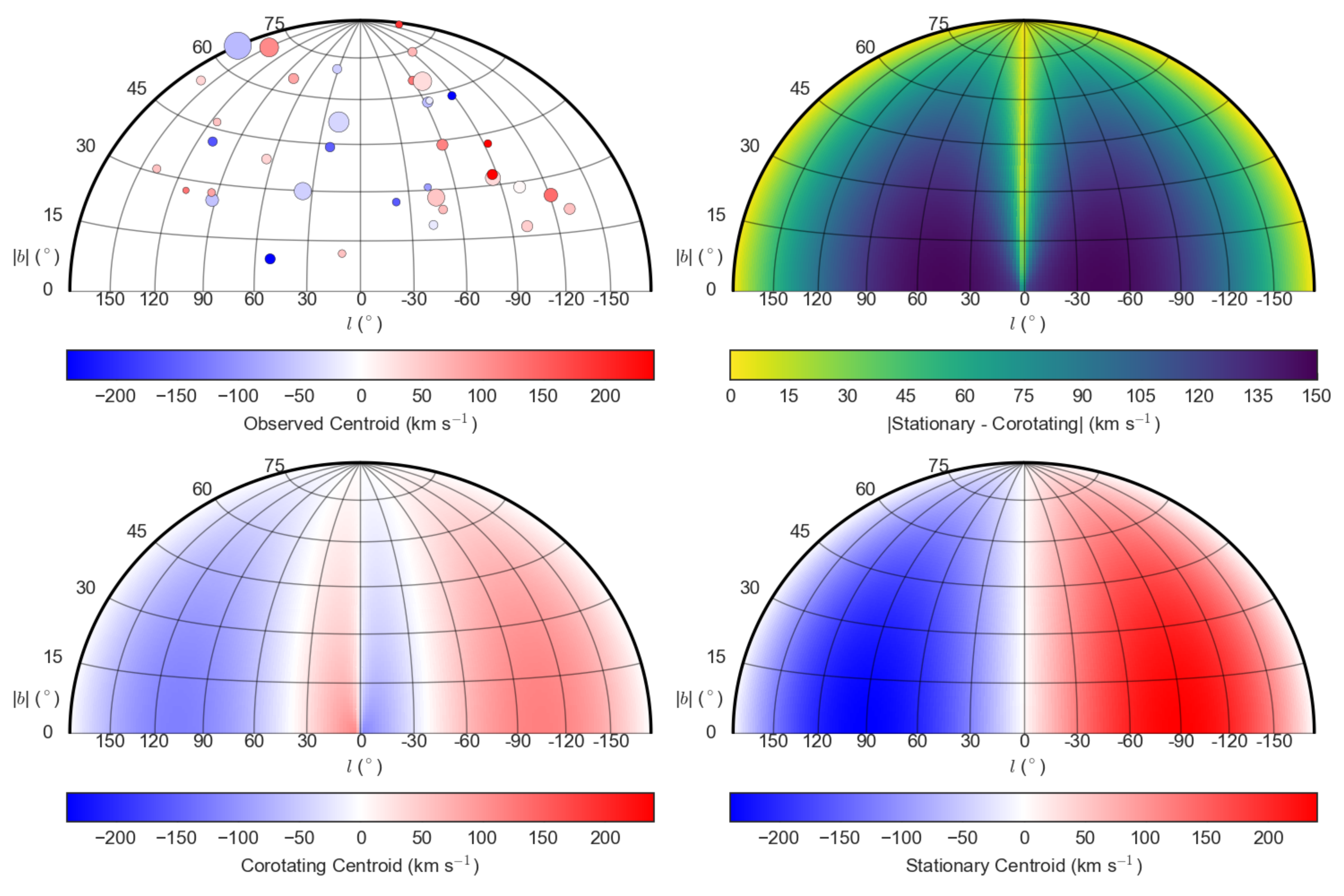}
\caption{Absorption line centroids in single-hemisphere Aitoff projections.  The upper-left panel shows the observed line centroid distribution from \cite{hodgeskl_etal16} where the point sizes are inversely proportional to their measurement errors.  The bottom panels represent model centroid calculations for a cororating (left) and stationary (right) halo gas distribution.  The upper-right panel shows the absolute difference between the stationary and corotating model, indicating where on the sky one has the greatest leverage to differentiate between models.
}
\label{figure.center_maps}
\end{figure*}


A useful approach for future analyses on these absorption lines will be the apparent optical depth method \citep{savage_sembach91}.  In this method, one decomposes the observed line profile into an apparent optical depth function $\tau_a(v)$ and converts this into an apparent column density function $N_a(v)$, which can be integrated to determine a total column density.  This method makes no assumption for the underlying line of sight density or line of sight velocity profile, but it would be useful to reconstruct a column density distribution similar to our Figure~\ref{figure.los_structure}.  The reason this method is not used in current analyses on the \ion{O}{7} absorbers is because: 1) the lines need to be at least partially resolved (FWHM(instrument) $\lesssim$ 2FWHM(line)) and 2) it requires multiple lines of the same species with different $f\lambda$ products to estimate saturation effects.  The lines are completely unresolved with current X-ray telescopes and detecting the \ion{O}{7} K$\beta$ transition at 18.63 \AA\ is difficult because it is intrinsically weaker than the K$\alpha$ transition and there is an RGS instrumental feature near the transition\footnote{http://xmm.esac.esa.int/external/xmm\_user\_support/ \\documentation/uhb/rgsmultipoint.html}.  Thus, this method will be useful if we obtain absorption line observations with a higher resolution spectrograph (see Section~\ref{subsection.future}).

\section{Discussion and Conclusions}
\label{section.discussion}

We have shown that the Milky Way's hot gas kinematics produce signatures in the intrinsic absorption line shapes and centroids.  Bulk velocity flows in the disk rotation direction along with a net inflow or outflow of gas can produce strong asymmetrical line shapes and significant line centroid deviations from the the absorption line rest wavelength.  These results alter the conversion between equivalent widths and column densities, producing significant deviations from both optically thin and zero velocity assumptions, but also imply the line shapes (not currently observable) and observed centroids encode information on the Milky Way's hot gas kinematics.  The new velocity considerations presented here motivate future analyses on the Milky Way's hot gas structure and kinematics.

\subsection{Implications for the Milky Way}
\label{subsection.applications}

The model line calculations presented in this paper form the basis for the \cite{hodgeskl_etal16} analysis on the Milky Way's hot gas kinematics.  Their non-parametric statistical tests concluded that the observed line centroid measurements were inconsistent with a stationary hot gas profile (a result not requiring additional modeling or assumptions for the gas distribution), whereas corotating model centroids from calculations in Section~\ref{subsection.line_profile} were more consistent with the data.  Figure~\ref{figure.center_maps} shows their observed centroid distribution on the sky with maps of the model line centroid distributions for the stationary and corotating profiles to illustrate how the models compare with the data.  They also used parametric modeling techniques with $v_{\phi}$ as a free parameter to show that a velocity profile lagging behind the disk with $v_{\phi}$ = 183 $\pm$ 41 km s$^{-1}$ is most consistent with the observed line centroid distribution.  The $v_z$ and $v_r$ flow parameters were also included, but the constraints were within 1$\sigma$ of 0 km s$^{-1}$.  These results are the first constraints on the Milky Way's $T \sim$10$^6$ K gas kinematics, and they depend on the absorption line profile calculations presented here.  The inferred velocities also have implications for other inferred hot gas properties, such as the density distribution and metallicity.

\begin{figure*}[!ht]
\centering
\includegraphics[width = 1.\textwidth]{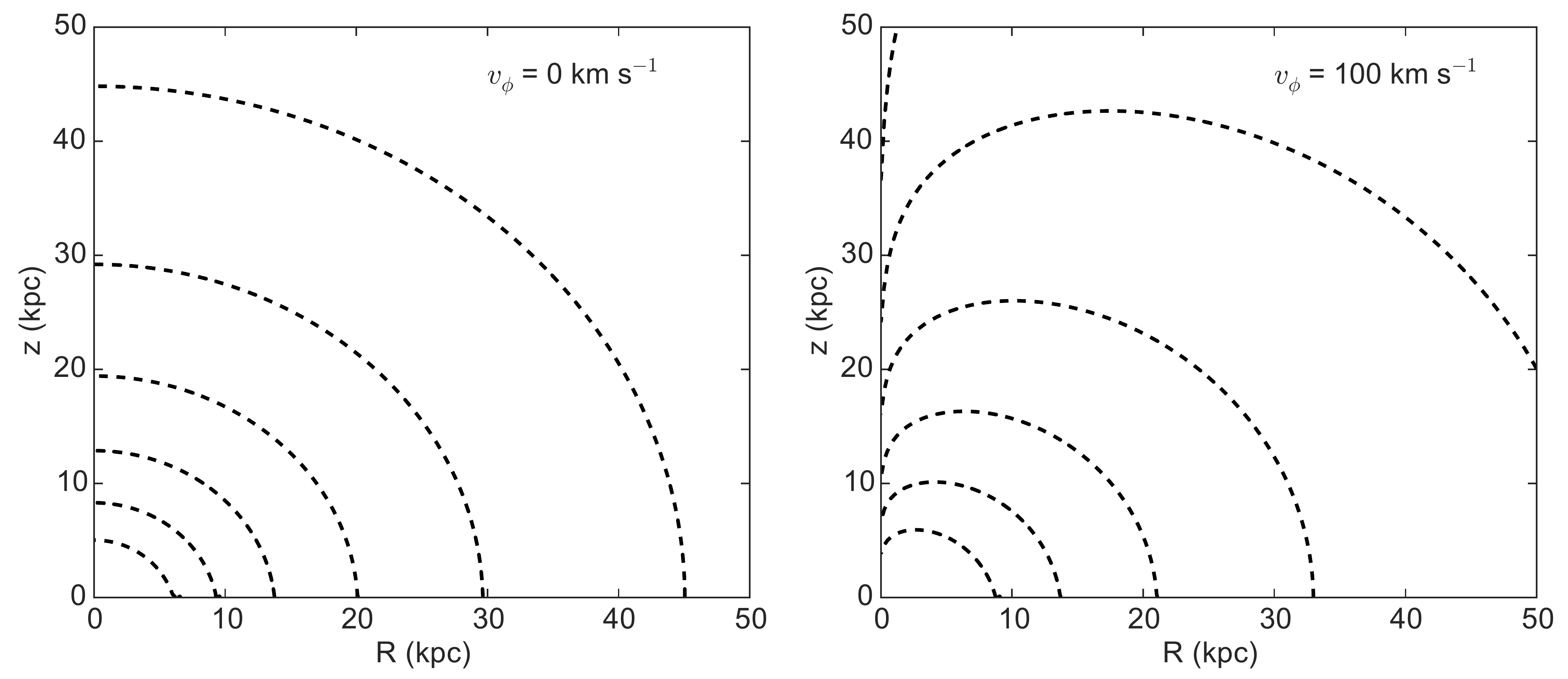}
\caption{Isodensity contours for stationary hot gas distribution (left) and a distribution with a flat 100 km s$^{-1}$ rotation curve (right).  The contours are normalized to the maximum density values, and are log-spaced with a 0.5 interval between -3 through -5.5.  Including rotation leads to a characteristic flattened morphology.  
}
\label{figure.flat_rot}
\end{figure*}


Significant hot gas rotation velocities can change the net potential well affecting the hot gas, resulting in a flattened or ``bow-tie'' morphology.  The prediction of hot, extended coronae around late-type galaxies is based on the assumption that the shock-heated gas exists in hydrostatic equilibrium with the dark matter potential well.  If the dark matter distribution is approximately spherical on large galactic scales and the gas is isothermal, hydrostatic equilibrium implies a spherical power law distribution for the hot gas:

	\begin{equation}
	\label{eq.hydrostatic}
    n \propto \text{exp} \left (-\frac{\Phi}{c_T^2} \right), 
	\end{equation}

\noindent where $\Phi$ is the potential ($\Phi$ $\propto$ $ln(r)$ for an isothermal sphere) and $c_T^2 = 2kT/\mu m_p$.  Rotation modifies the potential in the following way:

	\begin{equation}
	\label{eq.rotation}
    \Phi_{\text{eff}} = \Phi - \int^{R}_{R_o} \Omega^2(R')R'dR',
	\end{equation}

\noindent where $R_o$ is an arbitrary radius for normalization, $\Omega (R)$ is the angular rotation profile, and $\Phi_\text{eff}$ is the effective potential \citep[see Equation (8) in][]{barnabe_etal06}.  Inserting the effective potential into Equation~\ref{eq.hydrostatic} for flat rotation curves can alter the initial spherical profile into a flattened morphology.

To visualize this effect, we calculate an effective potential profile for a static potential and flat rotation curve we impose on the gas.  We assume the static potential includes a dark matter NFW profile \citep{nfw97} with a scale radius of 12 kpc and virial mass of $1.8 \times 10^{12}$ $M_{\odot}$ and a thin stellar disk profile with scale radius of 3.2 kpc and total surface density of 536 $M_{\odot}$ pc$^{\-2}$ \citep{binney_tremaine08}.  These assumptions do not account for the triaxiality of the Milky Way's dark matter distribution \citep{loebman_etal14} and the vertical structure of the stellar disk, but these calculations are meant to show the characteristic effect that rotation has on the hot gas.  We also point out that the baryonic components contribute very little to the potential for $r \gtrsim 5$ kpc, implying the details of their assumed profile do not significantly affect our calculations.  The results of these calculations for a stationary halo and a modest flat rotation curve of $v_{\phi}$ = 100 km s$^{-1}$ are seen in Figure~\ref{figure.flat_rot}.  One sees that the isodensity contours are constant with $r$ for the stationary halo, but become flattened with the inclusion of rotation.  We also point out that these calculations and general results are consistent with those found by \cite{marinacci_etal11}, who estimated these same effects of coronal rotation on the underlying morphology.

In contrast with these model predictions, several observation-based lines of evidence suggest the Milky Way's hot gas distribution is not significantly flattened.  As discussed in Section~\ref{subsection.model_assumptions}, analyses on large samples of line strength measurements, pulsar dispersion measures, and X-ray surface brightness measurements suggest a spherical morphology can better reproduce these observables than a disk-like morphology.  In particular, the models considered in \cite{miller_bregman13, miller_bregman15} included a flattened, extended distribution.  However, the model fitting results for this flattened model (with an additional parameter and thus one less degree of freedom) were not a significant improvement in fitting the data compared to a spherical model.  The analysis showed that a flattened halo was not required to fit the line strength measurements.  Also, observations of hot gas in other galaxies do not indicate rotation significantly impacts their morphologies.  The examples discussed in Section~\ref{subsection.model_assumptions} found that the extended X-ray halos around both early- and late-type galaxies are typically fit well with spherical $\beta$ models.  Additional analyses on higher-resolution X-ray images of early-type galaxies indicate any X-ray observed ellipticity does not correlate with the galaxies' rotation velocities, implying rotation does not drive their morphologies \citep{diehl_statler07}.  Thus, the balance of observational evidence suggests the extended hot gas distributions around the Milky Way and other galaxies is approximately spherical.

The observed line centroids from \cite{hodgeskl_etal16} combined with the models outlined in Section~\ref{subsection.line_profile} lead to the robust result that there is global kinematic structure in the hot gas, regardless of the assumed density profile.  This appears to be a valuable first step in this type of analysis, especially since there is little observational or theoretical background for the expected kinematic structure of this gas phase.  The line profiles presented here are designed to highlight how we can learn about the Milky Way's hot gas kinematics, and show how the hot gas kinematics can potentially impact the inferred hot gas structure.

One example of how these line profile calculations can improve constraints on the hot gas structure is with a more accurate conversion between measured equivalent width and the column density.  The velocity calculations presented in Section~\ref{subsection.equivalent_widths} imply that velocity flows modify the curve of growth analysis depending on the observed direction.  Thus, one needs to account for bulk velocity flows to infer an accurate column density from a measured equivalent width.  These effects have not been considered in previous analyses of the Milky Way's hot gas, and impact inferred hot gas properties.

As an example application of the improved equivalent width-column density conversion, we examine the hot gas metallicity based on the observed \ion{O}{7} equivalent width toward the LMC direction and the hot gas dispersion measure in this direction.  We convert the observed \ion{O}{7} equivalent width for the LMC X-3 X-ray binary into \ion{O}{7} column densities for different $b$ values with and without velocity effects.  These column densities are then converted to dispersion measures, which depend on the gas metallicity, and compared to the observed residual hot gas dispersion measure toward the LMC direction.  The dispersion measure is the integral of free electrons along the line of sight, $\int_{0}^{s} n(s) ds$, which relates to the \ion{O}{7} column density as follows:

\begin{figure}
\centering
\includegraphics[width = .45\textwidth]{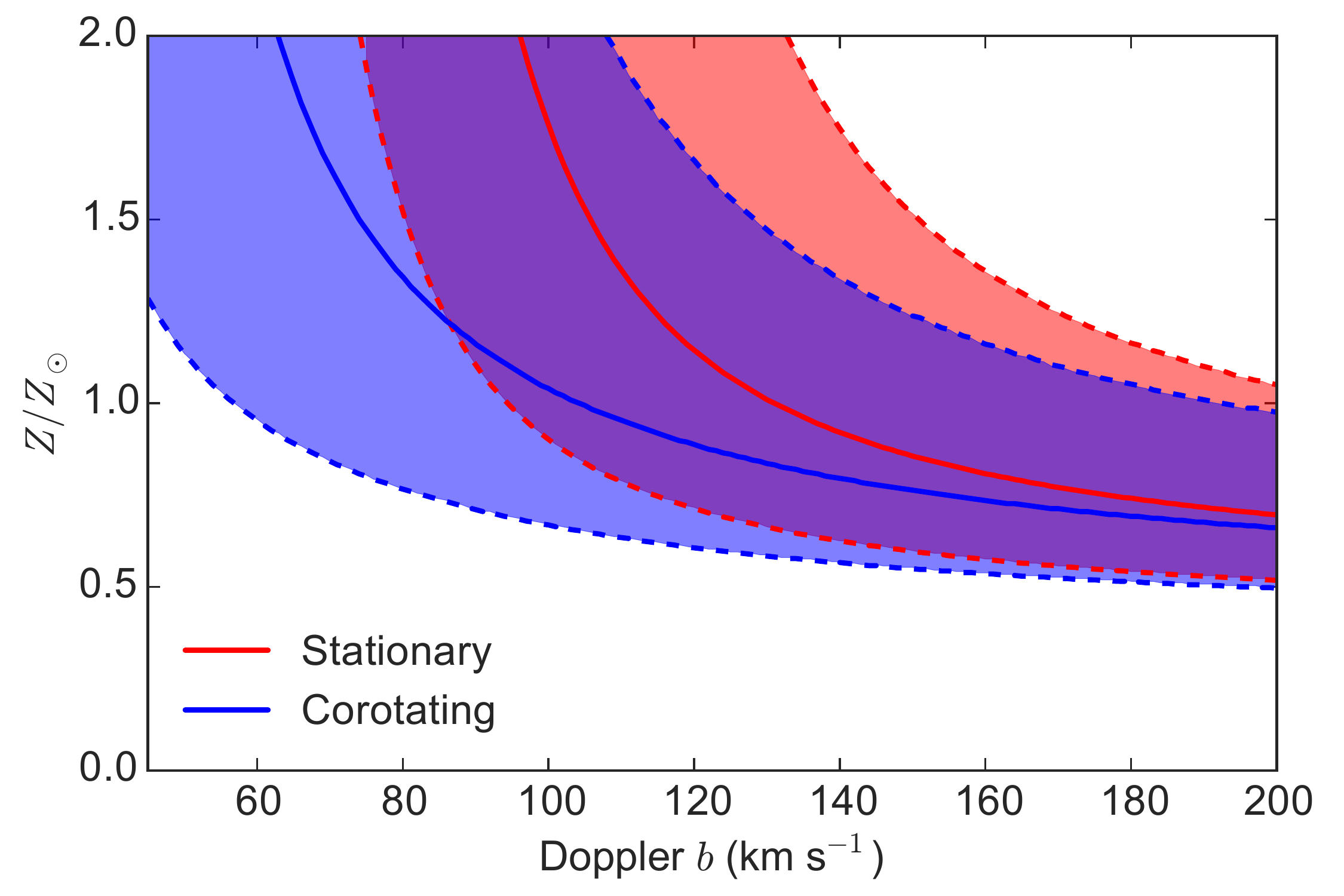}
\caption{The hot gas metallicity and Doppler $b$ values required for the observed LMC X-3 \ion{O}{7} equivalent width to equal a dispersion measure of 23 cm$^{-3}$ pc (solid lines).  The dashed lines incorporate the 90\% uncertainty on the measured equivalent width.  The red shaded region assumes a stationary halo for the column density conversion (or a standard curve of growth) while the blue shaded region assumes a corotating velocity profile.  
}
\label{figure.lmc_dm}
\end{figure}

    \begin{equation}
	\begin{aligned}
	 \label{eq.dm}
	  \text{DM} &= 1.2 \frac{N_{OVII}}{f_{OVII} A_{O} Z} \\
	  &= 14.2 \left( \frac{N_{OVII}}{10^{16} \text{cm}^{-2}} \right) 
	  \left(\frac{f_{OVII}}{0.5} \right)^{-1} \\
	  &\times \left(\frac{A_{O}}{5.49 \times 10^{-4}} \right)^{-1}
	  \left(\frac{Z}{Z_{\odot}} \right)^{-1} 
	  \text{cm}^{-3} \text{pc},
	\end{aligned}
    \end{equation}
    
\noindent where $f_{OVII}$ is the \ion{O}{7} ion fraction, $A_O$ is the fixed solar oxygen abundance relative to hydrogen \citep{holweger_01}, and $Z$ is the gas metallicity.  Thus, we infer an average hot gas metallicity by equating the inferred dispersion measure from the \ion{O}{7} measurement to the independently observed value.

The inferred metallicity depends on an accurate conversion between the observed \ion{O}{7} equivalent width and true column density, which we calculate above, and the residual dispersion measure due to hot gas.  For the latter, analyses on LMC pulsars suggest the Milky Way's hot gas contributes 23 cm$^{-3}$ pc to the total observed dispersion measures \citep{anderson_bregman10, fang_etal13}.  Independent analyses report a consistent \ion{O}{7} equivalent width for the LMC X-3 spectrum, with \cite{wang_etal05} suggesting a value of $20 \pm 6$ m\AA\ and \cite{bregman_ld07} suggesting a value of $21 \pm 5$ m\AA\ (90\% confidence regions).  Here, we use a more current measurement with improved calibration from the \cite{hodgeskl_etal16} analysis, $22^{+6}_{-4}$ m\AA.  We also assume all of the absorption arises from the Milky Way's hot gas distribution, as opposed to a distribution associated with the LMC or intrinsic to the LMC X-3 X-ray binary.  This appears to be a valid assumption since the \ion{O}{7} centroid is inconsistent with the LMC systemic velocity ($\sim$300 km s$^{-1}$) or the X-ray binary escape velocity ($\sim$10$^3$ km s$^{-1}$) \citep{wang_etal05, yao_etal09_b}.  We convert this equivalent width to a column density using our curves of growth for the LMC direction, a range of plausible $b$ values, and assuming either a stationary or corotating halo.  We then use Equation~\ref{eq.dm} to convert these to dispersion measures for different metallicities.  Figure~\ref{figure.lmc_dm} shows the lines where the calculated dispersion measure equals 23 cm$^{-3}$ pc.

\begin{figure*}[!ht]
\centering
\includegraphics[width = 1.\textwidth, keepaspectratio=true]{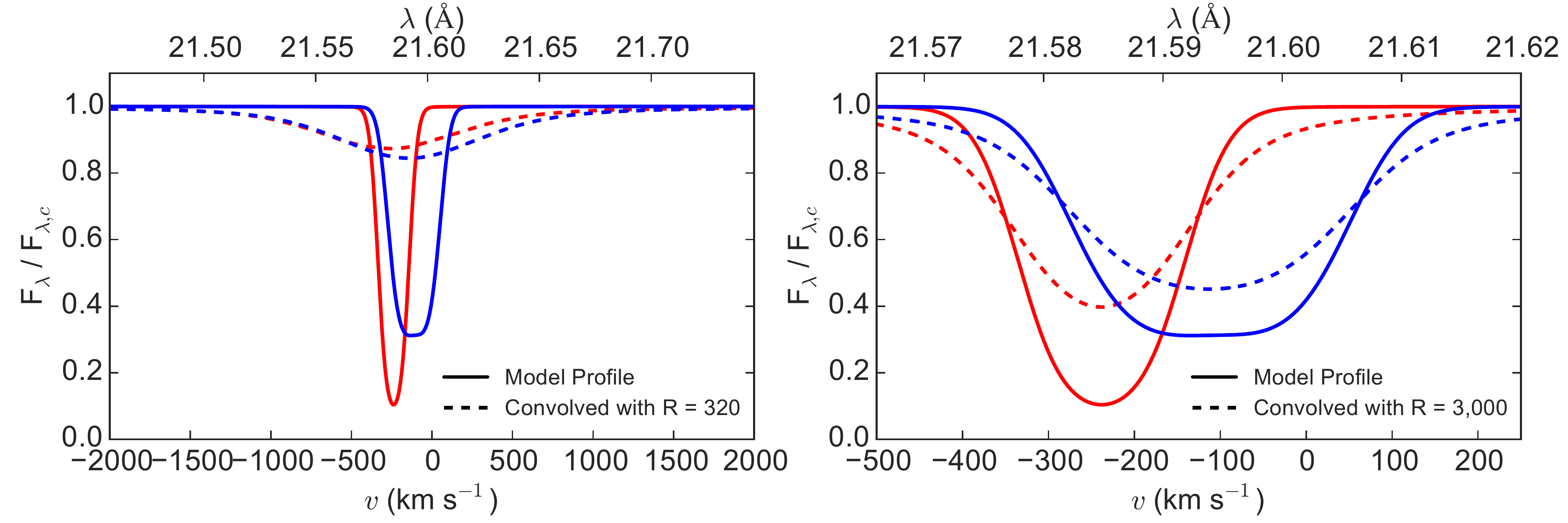}
\caption{Same model absorption line profiles as Figure~\ref{figure.line_detail} (solid lines), but including convolutions with current and future instrumental line spread functions (dashed lines).  The \textit{left} panel shows the model lines convolved with an \textit{XMM-Newton} RGS line spread function (approximately a Lorentzian with $R \approx$320), while the \textit{right} panel shows convolutions with an $R$ = 3000 Lorenztian profile.  The comparatively low resolution of current spectrographs only allows the equivalent widths and line centroids to be measured, but future high-resolution spectrographs will begin to resolve the line shapes and broadening due to velocity effects.
}
\label{figure.resolution}
\end{figure*}


There are several key results to note from the calculations represented in Figure~\ref{figure.lmc_dm}.  Increasing the $b$ value decreases the \ion{O}{7} column density required to match the observed equivalent width.  Therefore, increasing the $b$ value requires a decrease in metallicity in order to match the observed dispersion measure.  For a stationary halo, the transition between solar and sub-solar metallicity occurs at $b$ = 130 km s$^{-1}$.  Including velocity effects broadens the line profile in this direction, which further reduces the \ion{O}{7} column density required to match the observed equivalent width.  The transition between solar and sub-solar metallicity is at 100 km s$^{-1}$ in this case.  For both the stationary and corotating cases, the metallicity approaches a limit of $\approx$0.6$Z_{\odot}$ as $b$ increases beyond 200 km s$^{-1}$.  These results imply that the gas metallicity should be $\gtrsim$0.6$Z_{\odot}$ if $b$ is less than the sound speed, but also that $b \gtrsim $ 100 km s$^{-1}$ if the gas metallicity is not greater than solar.  This analysis provides important constraints on the average hot gas metallicity along the LMC sight line, however it also illustrates how the halo gas kinematics can impact inferred hot gas properties.

\subsection{Future Applications}
\label{subsection.future}

These velocity considerations have important applications for future analyses on the Milky Way's hot gas structure.  We have discussed several problems and initial attempts to solve them, such as the Milky Way's hot gas kinematic structure, the relation between the hot gas kinematics and density structure, and how to accurately interpret current hot gas observables.  Here, we discuss strategies to solve these issues in the future, and highlight how our velocity calculations will be useful.

We can better constrain the Milky Way's hot gas kinematic structure by targeting sight lines that predict the largest differences between model line centroid values.  These regions provide the most leverage when validating or excluding different types of velocity flows.  As an example, the upper-right panel in Figure~\ref{figure.center_maps} shows the absolute difference between the predicted line centroids for a stationary and corotating halo.  Clearly, the regions near $l$ = 90$\arcdeg$ and 270$\arcdeg$ show the largest differences of $\gtrsim$100 km s$^{-1}$, while regions near the Galactic pole and anticenter have differences of $\approx$0 km s$^{-1}$.  Although we do not show a map with a $v_r$ or $v_z$ flow, the Galactic pole and anticenter are ideal locations to probe those effects since rotation produces a negligible shift in these regions.  We also point out that a rotating gas profile predicts a distinctive line centroid feature near the Galactic center, in that the centroid switches sign for  $|l| \lesssim$ 30$\arcdeg$.  It is possible that this effect could be observed, although the Galactic center is a complex region with other kinematic features due to the Fermi bubbles \citep{su_etal10, fox_etal15}.  Regardless of these complications, these models should be used to motivate the most informative observations.

Interpreting current and future absorption line observations requires a better estimate for the plasma optical depth, or $b$ value.  Figure~\ref{figure.cog_corr} indicates optical depth effects and line of sight kinematics introduce significant corrections to the equivalent width-column density conversion.  The plasma $b$ value is usually determined by the \ion{O}{7} K$\beta$ to K$\alpha$ ratio, but weak K$\beta$ detections introduce systematic uncertainties in the inferred hot gas structural parameters \citep{williams_etal05, gupta_etal12, fang_etal15}.  One way to progress through this issue in the near future is with more sophisticated radiative transfer models.  Monte Carlo radiative transfer methods \citep{whitney11} are a promising avenue for these improvements since the codes predict both emission and absorption line strengths for an assumed plasma density profile and $b$ value.  Typical radiative transfer codes are not designed for analyses on the Milky Way's hot gas distribution, specifically with its extended geometry and our proximity embedded inside the distribution.  We are currently developing a special radiative transfer code accounting for these geometric effects and the gas kinematic structure to model the plasma radiation field and resultant line strengths.  Future projects with this code involve joint fits of the line strengths and shapes, resulting in improved estimates on the hot gas optical depth, structure, and kinematics.  These techniques allow us to maximize the information we have from current X-ray observations, but we can gain significantly more information from future X-ray telescopes.

An X-ray spectrograph with higher spectral resolution than current X-ray spectrographs will yield significantly improved constraints on the Milky Way's hot gas structure.  The \textit{XMM-Newton} RGS and future missions, such as \textit{Athena} and \textit{Astro-H}, cannot resolve the absorption lines with spectral resolutions of $E/ \Delta E \approx$300 - 400 near 0.6 keV \citep{denherder_etal03, denherder_etal12}.  The intrinsic line widths and deviations from Gaussian line shapes are sensitive to how $v_{\phi}$, $v_{r}$, and $v_{z}$ depend on $R$, $r$, $z$, implying these spectrographs cannot probe these signatures.  The left panel of Figure~\ref{figure.resolution} shows the stationary and corotating line profiles from Figure~\ref{figure.line_detail} convolved with a model RGS line spread function ($R \approx$ 320), where any line asymmetry or velocity broadening due to kinematic effects is lost due to the low resolution (the difference in line widths is only $\approx$20 km s$^{-1}$ in this case).  Alternatively, technology exists for X-ray grating spectrographs with $R \approx$3000 in the soft X-ray band \citep{smith_etal14b}.  At this spectral resolution, kinematic signatures, such as the velocity broadening and line centroid, are much clearer and easier to detect compared to current absorption line observations.  The right panel of Figure~\ref{figure.resolution} illustrates this improvement, where we convolve the same intrinsic line profiles discussed above with an $R$ = 3000 Lorentzian line spread function (the RGS line spread function is close to, but not a perfect Lorentzian).  Thus, an improved X-ray spectrograph will provide significantly more information on the hot gas kinematics than current instruments.

Given these improved observations and tools to interpret them, we will continue to develop more sophisticated velocity models motivated by galaxy formation predictions.  Galaxy formation simulations of Milky Way-like galaxies predict a diverse range of hot gas kinematic structures depending on the assumed formation mechanism.  The supernova-driven outflow scenario predicts hot gas rotational velocities comparable to the disk and $v_{r,z} \gtrsim$100 km s$^{-1}$ \citep[e.g., ][]{joung_maclow06}.  On the other hand, the shock-heated coronal gas scenario predicts a largely hydrostatic halo, but with significant rotational velocities ($\gtrsim$100 km s$^{-1}$) near the disk \citep[e.g., ][]{nuza_etal14}.  Additional complexities include the assumed stellar feedback prescription, the impact of galaxy-galaxy interactions, and the inferred coupling between the hot gas and cooler gas phases.  For example, \cite{marinacci_etal11} developed explored the mixing and angular momentum transfer between $\sim 10^4$ K galactic fountain gas ejected from the disk and an initially static $\sim 10^6$ K corona.  They found the cooler galactic fountain gas can accelerate the hotter coronal gas to velocities 80-120 km s$^{-1}$ slower than the disk, broadly consistent with the results from \cite{hodgeskl_etal16}.  Regardless of this consistency, there still exists many different velocity models that produce distinct absorption line signatures.  Future progress involves an iterative approach where velocity models become more complex to reflect a more realistic scenario, but the improved observations continue to validate or exclude different velocity models predicted in simulations.  In this way, theory and observations provide additional information how and when the hot gas obtains its kinematic structure.

\newpage
\subsection{Concluding Remarks}
\label{subsection.concluding}

The model absorption line results presented here have numerous important applications regarding the Milky Way's hot gas structure.  Many of the calculations were simple ideas, but were never applied to the Milky Way's extended hot gas distribution.  We applied simple kinematic models to absorption line calculations to show what can be observed now, how the gas kinematics can alter our interpretation of current observables, what can be observed in the future, and future strategies for exploring these effects.  Here, we summarize these results:

\begin{enumerate}

\item{The model line centroid calculations presented here combined with the observed line centroids from \cite{hodgeskl_etal16} indicate the Milky Way's hot gas distribution is likely not stationary.  A model velocity profile with significant rotation velocity ($\approx$180 km s$^{-1}$) is most consistent with the data.}

\item{Ordered velocity flows of this magnitude can impact constraints on the hot gas density structure.  Line of sight velocity effects alter the conversion between the equivalent width and column density.  The corrections from an optically thin column can range between a factor of $\approx$2 to $\approx$5 depending on the underlying kinematic structure.}
\item{These flows can alter the underlying density profile into a flattened or ``bow-tie'' morphology as discussed in \cite{marinacci_etal11}.  Current observational evidence suggests the Milky Way's hot gas is approximately spherical, although the data cannot conclusively rule out a flattened morphology.}

\item{Future observational strategies should focus on the regions with the largest difference between predicted model centroid values.  For probing the rotation curve, this is near the $l$ = 90$\arcdeg$ and 270$\arcdeg$ directions.  For probing accretion or outflows, this is near the Galactic pole or anticenter.  }

\item{Current kinematic constraints are limited by the resolution of X-ray spectrographs.  We show that an X-ray spectrograph with resolving power of $R \gtrsim $3000 would begin to resolve the intrinsic line profiles, resulting in significantly better constraints on the hot gas kinematics.}

\item{Future model line profile calculations should incorporate more detailed kinematic structure derived from galaxy formation simulations, and account for optical depth effects in the plasma with radiative transfer codes.  Both of these factors play significant roles in the absorption and emission line strength interpretations.}

\end{enumerate}

The work presented here can be considered an initial step toward understanding how these extended hot halos around late-type galaxies dynamically evolve with their host galaxies.


\acknowledgments

The authors thank Eric Bell for his helpful discussions on this project.  We also thank the anonymous referee for their helpful comments on the paper.  This research was funded by NASA Award NNX11AJ55G, and we thank this program for supporting this work.


\def\apjl{{ApJL}}               
\def\bain{{BAN}}               
\bibliographystyle{apj}

\begin{thebibliography}{}
\expandafter\ifx\csname natexlab\endcsname\relax\def\natexlab#1{#1}\fi

\bibitem[{{Anderson} \& {Bregman}(2010)}]{anderson_bregman10}
{Anderson}, M.~E., \& {Bregman}, J.~N. 2010, \apj, 714, 320

\bibitem[{{Anderson} \& {Bregman}(2011)}]{anderson_bregman11}
---. 2011, \apj, 737, 22

\bibitem[{{Anderson} {et~al.}(2016){Anderson}, {Churazov}, \&
  {Bregman}}]{anderson_etal16}
{Anderson}, M.~E., {Churazov}, E., \& {Bregman}, J.~N. 2016, \mnras, 455, 227

\bibitem[{{Barnab{\`e}} {et~al.}(2006){Barnab{\`e}}, {Ciotti}, {Fraternali}, \&
  {Sancisi}}]{barnabe_etal06}
{Barnab{\`e}}, M., {Ciotti}, L., {Fraternali}, F., \& {Sancisi}, R. 2006, \aap,
  446, 61

\bibitem[{{Bernardi} {et~al.}(2005){Bernardi}, {Sheth}, {Nichol}, {Schneider},
  \& {Brinkmann}}]{bernardi_etal05}
{Bernardi}, M., {Sheth}, R.~K., {Nichol}, R.~C., {Schneider}, D.~P., \&
  {Brinkmann}, J. 2005, \aj, 129, 61

\bibitem[{{Binney} \& {Tremaine}(2008)}]{binney_tremaine08}
{Binney}, J., \& {Tremaine}, S. 2008, {Galactic Dynamics: Second Edition}
  (Princeton University Press)

\bibitem[{{Bogd{\'a}n} {et~al.}(2013{\natexlab{a}}){Bogd{\'a}n}, {Forman},
  {Kraft}, \& {Jones}}]{bogdan_etal13b}
{Bogd{\'a}n}, {\'A}., {Forman}, W.~R., {Kraft}, R.~P., \& {Jones}, C.
  2013{\natexlab{a}}, \apj, 772, 98

\bibitem[{{Bogd{\'a}n} {et~al.}(2013{\natexlab{b}}){Bogd{\'a}n}, {Forman},
  {Vogelsberger}, {Bourdin}, {Sijacki}, {Mazzotta}, {Kraft}, {Jones},
  {Gilfanov}, {Churazov}, \& {David}}]{bogdan_etal13a}
{Bogd{\'a}n}, {\'A}., {Forman}, W.~R., {Vogelsberger}, M., {et~al.}
  2013{\natexlab{b}}, \apj, 772, 97

\bibitem[{{Bowen} {et~al.}(2008){Bowen}, {Jenkins}, {Tripp}, {Sembach},
  {Savage}, {Moos}, {Oegerle}, {Friedman}, {Gry}, {Kruk}, {Murphy}, {Sankrit},
  {Shull}, {Sonneborn}, \& {York}}]{bowen_etal08}
{Bowen}, D.~V., {Jenkins}, E.~B., {Tripp}, T.~M., {et~al.} 2008, \apjs, 176, 59

\bibitem[{{Boylan-Kolchin} {et~al.}(2013){Boylan-Kolchin}, {Bullock}, {Sohn},
  {Besla}, \& {van der Marel}}]{boylan_kolchin_etal13}
{Boylan-Kolchin}, M., {Bullock}, J.~S., {Sohn}, S.~T., {Besla}, G., \& {van der
  Marel}, R.~P. 2013, \apj, 768, 140

\bibitem[{{Bregman} {et~al.}(2015){Bregman}, {Alves}, {Miller}, \&
  {Hodges-Kluck}}]{bregman_etal15}
{Bregman}, J.~N., {Alves}, G.~C., {Miller}, M.~J., \& {Hodges-Kluck}, E. 2015,
  Journal of Astronomical Telescopes, Instruments, and Systems, 1, 045003

\bibitem[{{Bregman} \& {Lloyd-Davies}(2007)}]{bregman_ld07}
{Bregman}, J.~N., \& {Lloyd-Davies}, E.~J. 2007, \apj, 669, 990

\bibitem[{{Burton}(1976)}]{burton76}
{Burton}, W.~B. 1976, \araa, 14, 275

\bibitem[{{Carter} \& {Sembay}(2008)}]{carter_sembay08}
{Carter}, J.~A., \& {Sembay}, S. 2008, \aap, 489, 837

\bibitem[{{Carter} {et~al.}(2011){Carter}, {Sembay}, \& {Read}}]{carter_etal11}
{Carter}, J.~A., {Sembay}, S., \& {Read}, A.~M. 2011, \aap, 527, A115

\bibitem[{{Cen}(2012)}]{cen12}
{Cen}, R. 2012, \apj, 753, 17

\bibitem[{{Cen} \& {Ostriker}(2006)}]{cen_ostriker06}
{Cen}, R., \& {Ostriker}, J.~P. 2006, \apj, 650, 560

\bibitem[{{Coil} {et~al.}(2011){Coil}, {Weiner}, {Holz}, {Cooper}, {Yan}, \&
  {Aird}}]{coil_etal11}
{Coil}, A.~L., {Weiner}, B.~J., {Holz}, D.~E., {et~al.} 2011, \apj, 743, 46

\bibitem[{{Dai} {et~al.}(2012){Dai}, {Anderson}, {Bregman}, \&
  {Miller}}]{dai_etal12}
{Dai}, X., {Anderson}, M.~E., {Bregman}, J.~N., \& {Miller}, J.~M. 2012, \apj,
  755, 107

\bibitem[{{de Blok} {et~al.}(2008){de Blok}, {Walter}, {Brinks},
  {Trachternach}, {Oh}, \& {Kennicutt}}]{deblok_etal08}
{de Blok}, W.~J.~G., {Walter}, F., {Brinks}, E., {et~al.} 2008, \aj, 136, 2648

\bibitem[{{den Herder} {et~al.}(2012){den Herder}, {Bagnali}, {Bandler},
  {Barbera}, {Barcons}, {Barret}, {Bastia}, {Bisotti}, {Boyce}, {Cara},
  {Ceballos}, {Corcione}, {Cobo}, {Colasanti}, {de Plaa}, {DiPirro}, {Doriese},
  {Ezoe}, {Fujimoto}, {Gatti}, {Gottardi}, {Guttridge}, {den Hartog},
  {Hepburn}, {Kelley}, {Irwin}, {Ishisaki}, {Kilbourne}, {de Korte}, {van der
  Kuur}, {Lotti}, {Macculi}, {Mitsuda}, {Mineo}, {Natalucci}, {Ohashi}, {Page},
  {Paltani}, {Perinati}, {Piro}, {Pigot}, {Porter}, {Rauw}, {Ravera},
  {Renotte}, {Sauvageot}, {Schmid}, {Sciortino}, {Shirron}, {Takei},
  {Torrioli}, {Tsujimoto}, {Valenziano}, {Willingale}, {de Vries}, {van Weers},
  {Wilms}, \& {Yamasaki}}]{denherder_etal12}
{den Herder}, J.~W., {Bagnali}, D., {Bandler}, S., {et~al.} 2012, \procspie,
  8443, 2

\bibitem[{{den Herder} {et~al.}(2003){den Herder}, {Brinkman}, {Kahn},
  {Branduardi-Raymont}, {Audard}, {Behar}, {Blustin}, {den Boggende}, {Cottam},
  {Erd}, {Gabriel}, {Guedel}, {van der Heyden}, {Kaastra}, {Kinkhabwala},
  {Leutenegger}, {Mewe}, {Paerels}, {Raassen}, {Peterson}, {Pollock},
  {Rasmussen}, {Sako}, {Santos-Lleo}, {Steenbrugge}, {Tamura}, \& {de
  Vries}}]{denherder_etal03}
{den Herder}, J.-W.~W., {Brinkman}, A.~C., {Kahn}, S.~M., {et~al.} 2003,
  \procspie, 4851, 196

\bibitem[{{Diehl} \& {Statler}(2007)}]{diehl_statler07}
{Diehl}, S., \& {Statler}, T.~S. 2007, \apj, 668, 150

\bibitem[{{Fang} {et~al.}(2013){Fang}, {Bullock}, \&
  {Boylan-Kolchin}}]{fang_etal13}
{Fang}, T., {Bullock}, J., \& {Boylan-Kolchin}, M. 2013, \apj, 762, 20

\bibitem[{{Fang} {et~al.}(2015){Fang}, {Buote}, {Bullock}, \&
  {Ma}}]{fang_etal15}
{Fang}, T., {Buote}, D., {Bullock}, J., \& {Ma}, R. 2015, \apjs, 217, 21

\bibitem[{{Forman} {et~al.}(1985){Forman}, {Jones}, \&
  {Tucker}}]{forman_etal85}
{Forman}, W., {Jones}, C., \& {Tucker}, W. 1985, \apj, 293, 102

\bibitem[{{Fox} {et~al.}(2015){Fox}, {Bordoloi}, {Savage}, {Lockman},
  {Jenkins}, {Wakker}, {Bland-Hawthorn}, {Hernandez}, {Kim}, {Benjamin},
  {Bowen}, \& {Tumlinson}}]{fox_etal15}
{Fox}, A.~J., {Bordoloi}, R., {Savage}, B.~D., {et~al.} 2015, \apjl, 799, L7

\bibitem[{{Fukugita} \& {Peebles}(2006)}]{fukugita_peebles06}
{Fukugita}, M., \& {Peebles}, P.~J.~E. 2006, \apj, 639, 590

\bibitem[{{Galeazzi} {et~al.}(2014){Galeazzi}, {Chiao}, {Collier}, {Cravens},
  {Koutroumpa}, {Kuntz}, {Lallement}, {Lepri}, {McCammon}, {Morgan}, {Porter},
  {Robertson}, {Snowden}, {Thomas}, {Uprety}, {Ursino}, \&
  {Walsh}}]{galeazzi_etal14}
{Galeazzi}, M., {Chiao}, M., {Collier}, M.~R., {et~al.} 2014, \nat, 512, 171

\bibitem[{{Grcevich} \& {Putman}(2009)}]{grcevich_putman09}
{Grcevich}, J., \& {Putman}, M.~E. 2009, \apj, 696, 385

\bibitem[{{Gupta} {et~al.}(2012){Gupta}, {Mathur}, {Krongold}, {Nicastro}, \&
  {Galeazzi}}]{gupta_etal12}
{Gupta}, A., {Mathur}, S., {Krongold}, Y., {Nicastro}, F., \& {Galeazzi}, M.
  2012, \apjl, 756, L8

\bibitem[{{Hagihara} {et~al.}(2010){Hagihara}, {Yao}, {Yamasaki}, {Mitsuda},
  {Wang}, {Takei}, {Yoshino}, \& {McCammon}}]{hagihara_etal10}
{Hagihara}, T., {Yao}, Y., {Yamasaki}, N.~Y., {et~al.} 2010, \pasj, 62, 723

\bibitem[{{Henley} \& {Shelton}(2012)}]{hs12}
{Henley}, D.~B., \& {Shelton}, R.~L. 2012, \apjs, 202, 14

\bibitem[{{Henley} \& {Shelton}(2013)}]{hs13}
---. 2013, \apj, 773, 92

\bibitem[{{Henley} \& {Shelton}(2015)}]{hs15}
---. 2015, \apj, 808, 22

\bibitem[{{Hill} {et~al.}(2012){Hill}, {Joung}, {Mac Low}, {Benjamin},
  {Haffner}, {Klingenberg}, \& {Waagan}}]{hill_etal12}
{Hill}, A.~S., {Joung}, M.~R., {Mac Low}, M.-M., {et~al.} 2012, \apj, 750, 104

\bibitem[{{Hodges-Kluck} {et~al.}(2016){Hodges-Kluck}, {Miller}, \&
  {Bregman}}]{hodgeskl_etal16}
{Hodges-Kluck}, E., {Miller}, M.~J., \& {Bregman}, J.~N. 2016, \apjl, submitted

\bibitem[{{Holweger}(2001)}]{holweger_01}
{Holweger}, H. 2001, in American Institute of Physics Conference Series, Vol.
  598, Joint SOHO/ACE workshop ''Solar and Galactic Composition'', ed. R.~F.
  {Wimmer-Schweingruber}, 23--30

\bibitem[{{Joung} \& {Mac Low}(2006)}]{joung_maclow06}
{Joung}, M.~K.~R., \& {Mac Low}, M.-M. 2006, \apj, 653, 1266

\bibitem[{{Kacprzak} {et~al.}(2012){Kacprzak}, {Churchill}, \&
  {Nielsen}}]{kacprzak_etal12}
{Kacprzak}, G.~G., {Churchill}, C.~W., \& {Nielsen}, N.~M. 2012, \apjl, 760, L7

\bibitem[{{Kalberla} {et~al.}(2007){Kalberla}, {Dedes}, {Kerp}, \&
  {Haud}}]{kalberla_etal07}
{Kalberla}, P.~M.~W., {Dedes}, L., {Kerp}, J., \& {Haud}, U. 2007, \aap, 469,
  511

\bibitem[{{Kaufmann} {et~al.}(2006){Kaufmann}, {Mayer}, {Wadsley}, {Stadel}, \&
  {Moore}}]{kaufmann_etal06}
{Kaufmann}, T., {Mayer}, L., {Wadsley}, J., {Stadel}, J., \& {Moore}, B. 2006,
  \mnras, 370, 1612

\bibitem[{{Kent}(1987)}]{kent87}
{Kent}, S.~M. 1987, \aj, 93, 816

\bibitem[{{Klypin} {et~al.}(2002){Klypin}, {Zhao}, \&
  {Somerville}}]{klypin_etal02}
{Klypin}, A., {Zhao}, H., \& {Somerville}, R.~S. 2002, \apj, 573, 597

\bibitem[{{Koutroumpa} {et~al.}(2007){Koutroumpa}, {Acero}, {Lallement},
  {Ballet}, \& {Kharchenko}}]{koutroumpa_etal07}
{Koutroumpa}, D., {Acero}, F., {Lallement}, R., {Ballet}, J., \& {Kharchenko},
  V. 2007, \aap, 475, 901

\bibitem[{{Kuntz} \& {Snowden}(2000)}]{kuntz_snowden00}
{Kuntz}, K.~D., \& {Snowden}, S.~L. 2000, \apj, 543, 195

\bibitem[{{Kwak} \& {Shelton}(2010)}]{kwak_shelton10}
{Kwak}, K., \& {Shelton}, R.~L. 2010, \apj, 719, 523

\bibitem[{{Lallement} {et~al.}(2003){Lallement}, {Welsh}, {Vergely}, {Crifo},
  \& {Sfeir}}]{lallement_etal03}
{Lallement}, R., {Welsh}, B.~Y., {Vergely}, J.~L., {Crifo}, F., \& {Sfeir}, D.
  2003, \aap, 411, 447

\bibitem[{{Loeb} {et~al.}(2005){Loeb}, {Reid}, {Brunthaler}, \&
  {Falcke}}]{loeb_etal05}
{Loeb}, A., {Reid}, M.~J., {Brunthaler}, A., \& {Falcke}, H. 2005, \apj, 633,
  894

\bibitem[{{Loebman} {et~al.}(2014){Loebman}, {Ivezi{\'c}}, {Quinn}, {Bovy},
  {Christensen}, {Juri{\'c}}, {Ro{\v s}kar}, {Brooks}, \&
  {Governato}}]{loebman_etal14}
{Loebman}, S.~R., {Ivezi{\'c}}, {\v Z}., {Quinn}, T.~R., {et~al.} 2014, \apj,
  794, 151

\bibitem[{{Marinacci} {et~al.}(2011){Marinacci}, {Fraternali}, {Nipoti},
  {Binney}, {Ciotti}, \& {Londrillo}}]{marinacci_etal11}
{Marinacci}, F., {Fraternali}, F., {Nipoti}, C., {et~al.} 2011, \mnras, 415,
  1534

\bibitem[{{Marinacci} {et~al.}(2014){Marinacci}, {Pakmor}, {Springel}, \&
  {Simpson}}]{marinacci_etal14}
{Marinacci}, F., {Pakmor}, R., {Springel}, V., \& {Simpson}, C.~M. 2014,
  \mnras, 442, 3745

\bibitem[{{Matkovi{\'c}} \& {Guzm{\'a}n}(2005)}]{matkovic_guzman05}
{Matkovi{\'c}}, A., \& {Guzm{\'a}n}, R. 2005, \mnras, 362, 289

\bibitem[{{McCammon} {et~al.}(2002){McCammon}, {Almy}, {Apodaca}, {Bergmann
  Tiest}, {Cui}, {Deiker}, {Galeazzi}, {Juda}, {Lesser}, {Mihara},
  {Morgenthaler}, {Sanders}, {Zhang}, {Figueroa-Feliciano}, {Kelley},
  {Moseley}, {Mushotzky}, {Porter}, {Stahle}, \&
  {Szymkowiak}}]{mccammon_etal02}
{McCammon}, D., {Almy}, R., {Apodaca}, E., {et~al.} 2002, \apj, 576, 188

\bibitem[{{Miller} \& {Bregman}(2013)}]{miller_bregman13}
{Miller}, M.~J., \& {Bregman}, J.~N. 2013, \apj, 770, 118

\bibitem[{{Miller} \& {Bregman}(2015)}]{miller_bregman15}
---. 2015, \apj, 800, 14

\bibitem[{{Navarro} {et~al.}(1997){Navarro}, {Frenk}, \& {White}}]{nfw97}
{Navarro}, J.~F., {Frenk}, C.~S., \& {White}, S.~D.~M. 1997, \apj, 490, 493

\bibitem[{{Nicastro} {et~al.}(2002){Nicastro}, {Zezas}, {Drake}, {Elvis},
  {Fiore}, {Fruscione}, {Marengo}, {Mathur}, \& {Bianchi}}]{nicastro_etal02}
{Nicastro}, F., {Zezas}, A., {Drake}, J., {et~al.} 2002, \apj, 573, 157

\bibitem[{{Nuza} {et~al.}(2014){Nuza}, {Parisi}, {Scannapieco}, {Richter},
  {Gottl{\"o}ber}, \& {Steinmetz}}]{nuza_etal14}
{Nuza}, S.~E., {Parisi}, F., {Scannapieco}, C., {et~al.} 2014, \mnras, 441,
  2593

\bibitem[{{O'Sullivan} {et~al.}(2003){O'Sullivan}, {Ponman}, \&
  {Collins}}]{osullivan_etal03}
{O'Sullivan}, E., {Ponman}, T.~J., \& {Collins}, R.~S. 2003, \mnras, 340, 1375

\bibitem[{{Paerels} \& {Kahn}(2003)}]{paerels_kahn03}
{Paerels}, F.~B.~S., \& {Kahn}, S.~M. 2003, \araa, 41, 291

\bibitem[{{Putman} {et~al.}(2012){Putman}, {Peek}, \& {Joung}}]{putman_etal12}
{Putman}, M.~E., {Peek}, J.~E.~G., \& {Joung}, M.~R. 2012, \araa, 50, 491

\bibitem[{{Rasmussen} {et~al.}(2003){Rasmussen}, {Kahn}, \&
  {Paerels}}]{rasmussen_etal03}
{Rasmussen}, A., {Kahn}, S.~M., \& {Paerels}, F. 2003, in Astrophysics and
  Space Science Library, Vol. 281, The IGM/Galaxy Connection. The Distribution
  of Baryons at z=0, ed. J.~L. {Rosenberg} \& M.~E. {Putman} (Dordrecht:
  Kluwer), 109

\bibitem[{{Reid} {et~al.}(2014){Reid}, {Menten}, {Brunthaler}, {Zheng}, {Dame},
  {Xu}, {Wu}, {Zhang}, {Sanna}, {Sato}, {Hachisuka}, {Choi}, {Immer},
  {Moscadelli}, {Rygl}, \& {Bartkiewicz}}]{reid_etal14}
{Reid}, M.~J., {Menten}, K.~M., {Brunthaler}, A., {et~al.} 2014, \apj, 783, 130

\bibitem[{{Savage} \& {Sembach}(1991)}]{savage_sembach91}
{Savage}, B.~D., \& {Sembach}, K.~R. 1991, \apj, 379, 245

\bibitem[{{Savage} {et~al.}(2003){Savage}, {Sembach}, {Wakker}, {Richter},
  {Meade}, {Jenkins}, {Shull}, {Moos}, \& {Sonneborn}}]{savage_etal03}
{Savage}, B.~D., {Sembach}, K.~R., {Wakker}, B.~P., {et~al.} 2003, \apjs, 146,
  125

\bibitem[{{Sembach} {et~al.}(2003){Sembach}, {Wakker}, {Savage}, {Richter},
  {Meade}, {Shull}, {Jenkins}, {Sonneborn}, \& {Moos}}]{sembach_etal03}
{Sembach}, K.~R., {Wakker}, B.~P., {Savage}, B.~D., {et~al.} 2003, \apjs, 146,
  165

\bibitem[{{Sembay} {et~al.}(2004){Sembay}, {Abbey}, {Altieri}, {Ambrosi},
  {Baskill}, {Ferrando}, {Mukerjee}, {Read}, \& {Turner}}]{sembay_etal04}
{Sembay}, S., {Abbey}, A., {Altieri}, B., {et~al.} 2004, \procspie, 5488, 264

\bibitem[{{Shattow} \& {Loeb}(2009)}]{shattow_loeb09}
{Shattow}, G., \& {Loeb}, A. 2009, \mnras, 392, L21

\bibitem[{{Smith} {et~al.}(2014{\natexlab{a}}){Smith}, {Foster}, {Edgar}, \&
  {Brickhouse}}]{smith_etal14}
{Smith}, R.~K., {Foster}, A.~R., {Edgar}, R.~J., \& {Brickhouse}, N.~S.
  2014{\natexlab{a}}, \apj, 787, 77

\bibitem[{{Smith} {et~al.}(2007){Smith}, {Bautz}, {Edgar}, {Fujimoto},
  {Hamaguchi}, {Hughes}, {Ishida}, {Kelley}, {Kilbourne}, {Kuntz}, {McCammon},
  {Miller}, {Mitsuda}, {Mukai}, {Plucinsky}, {Porter}, {Snowden}, {Takei},
  {Terada}, {Tsuboi}, \& {Yamasaki}}]{smith_etal07}
{Smith}, R.~K., {Bautz}, M.~W., {Edgar}, R.~J., {et~al.} 2007, \pasj, 59, 141

\bibitem[{{Smith} {et~al.}(2014{\natexlab{b}}){Smith}, {Ackermann}, {Allured},
  {Bautz}, {Bregman}, {Bookbinder}, {Burrows}, {Brenneman}, {Brickhouse},
  {Cheimets}, {Carrier}, {Freeman}, {Kaastra}, {McEntaffer}, {Miller}, {Ptak},
  {Petre}, \& {Vacanti}}]{smith_etal14b}
{Smith}, R.~K., {Ackermann}, M., {Allured}, R., {et~al.} 2014{\natexlab{b}},
  \procspie, 9144, 4

\bibitem[{{Snowden} {et~al.}(2004){Snowden}, {Collier}, \&
  {Kuntz}}]{snowden_etal04}
{Snowden}, S.~L., {Collier}, M.~R., \& {Kuntz}, K.~D. 2004, \apj, 610, 1182

\bibitem[{{Snowden} {et~al.}(1990){Snowden}, {Cox}, {McCammon}, \&
  {Sanders}}]{snowden_etal90}
{Snowden}, S.~L., {Cox}, D.~P., {McCammon}, D., \& {Sanders}, W.~T. 1990, \apj,
  354, 211

\bibitem[{{Snowden} {et~al.}(1993){Snowden}, {McCammon}, \&
  {Verter}}]{snowden_etal93}
{Snowden}, S.~L., {McCammon}, D., \& {Verter}, F. 1993, \apjl, 409, L21

\bibitem[{{Snowden} {et~al.}(1997){Snowden}, {Egger}, {Freyberg}, {McCammon},
  {Plucinsky}, {Sanders}, {Schmitt}, {Tr{\"u}mper}, \&
  {Voges}}]{snowden_etal97}
{Snowden}, S.~L., {Egger}, R., {Freyberg}, M.~J., {et~al.} 1997, \apj, 485, 125

\bibitem[{{Spitzer}(1956)}]{spitzer56}
{Spitzer}, Jr., L. 1956, \apj, 124, 20

\bibitem[{{Su} {et~al.}(2010){Su}, {Slatyer}, \& {Finkbeiner}}]{su_etal10}
{Su}, M., {Slatyer}, T.~R., \& {Finkbeiner}, D.~P. 2010, \apj, 724, 1044

\bibitem[{{Sutherland} \& {Dopita}(1993)}]{sutherland_dopita93}
{Sutherland}, R.~S., \& {Dopita}, M.~A. 1993, \apjs, 88, 253

\bibitem[{{Toft} {et~al.}(2002){Toft}, {Rasmussen}, {Sommer-Larsen}, \&
  {Pedersen}}]{toft_etal02}
{Toft}, S., {Rasmussen}, J., {Sommer-Larsen}, J., \& {Pedersen}, K. 2002,
  \mnras, 335, 799

\bibitem[{{Wakker} \& {van Woerden}(1997)}]{wakker_vanwoerden97}
{Wakker}, B.~P., \& {van Woerden}, H. 1997, \araa, 35, 217

\bibitem[{{Wang} {et~al.}(2005){Wang}, {Yao}, {Tripp}, {Fang}, {Cui},
  {Nicastro}, {Mathur}, {Williams}, {Song}, \& {Croft}}]{wang_etal05}
{Wang}, Q.~D., {Yao}, Y., {Tripp}, T.~M., {et~al.} 2005, \apj, 635, 386

\bibitem[{{Welsh} \& {Shelton}(2009)}]{welsh_shelton09}
{Welsh}, B.~Y., \& {Shelton}, R.~L. 2009, \apss, 323, 1

\bibitem[{{Westerhout}(1957)}]{westerhout57}
{Westerhout}, G. 1957, \bain, 13, 201

\bibitem[{{White} \& {Frenk}(1991)}]{white_frenk91}
{White}, S.~D.~M., \& {Frenk}, C.~S. 1991, \apj, 379, 52

\bibitem[{{Whitney}(2011)}]{whitney11}
{Whitney}, B.~A. 2011, BASI, 39, 101

\bibitem[{{Williams} {et~al.}(2005){Williams}, {Mathur}, {Nicastro}, {Elvis},
  {Drake}, {Fang}, {Fiore}, {Krongold}, {Wang}, \& {Yao}}]{williams_etal05}
{Williams}, R.~J., {Mathur}, S., {Nicastro}, F., {et~al.} 2005, \apj, 631, 856

\bibitem[{{Yao} \& {Wang}(2005)}]{yao_wang05}
{Yao}, Y., \& {Wang}, Q.~D. 2005, \apj, 624, 751

\bibitem[{{Yao} \& {Wang}(2007)}]{yao_wang07}
---. 2007, \apj, 658, 1088

\bibitem[{{Yao} {et~al.}(2009){Yao}, {Wang}, {Hagihara}, {Mitsuda}, {McCammon},
  \& {Yamasaki}}]{yao_etal09_b}
{Yao}, Y., {Wang}, Q.~D., {Hagihara}, T., {et~al.} 2009, \apj, 690, 143

\bibitem[{{Yoshino} {et~al.}(2009){Yoshino}, {Mitsuda}, {Yamasaki}, {Takei},
  {Hagihara}, {Masui}, {Bauer}, {McCammon}, {Fujimoto}, {Wang}, \&
  {Yao}}]{yoshino_etal09}
{Yoshino}, T., {Mitsuda}, K., {Yamasaki}, N.~Y., {et~al.} 2009, \pasj, 61, 805

\end{thebibliography}

\end{document}